\documentclass{article}
\usepackage{authblk}
\usepackage{geometry}
\usepackage[titletoc]{appendix}
\usepackage{enumitem}
\usepackage{booktabs, diagbox}
\usepackage{xcolor}
\usepackage{cancel}

\newcommand{\ad}{\mathrm{ad}\ }

\usepackage{tikz}
\usetikzlibrary{cd, arrows}
\usepackage{amssymb, amsmath, amsthm, mathtools, slashed, bm}
\def\-{\raisebox{.75pt}{-}}

\numberwithin{table}{section}
\numberwithin{figure}{section}
\numberwithin{equation}{section}

\theoremstyle{definition}
\newtheorem{defn}{Definition}[section]

\newtheorem{exmp}{Example}[section]
\newtheorem{rmk}{Remark}[section]

\theoremstyle{plain}
\newtheorem{lem}{Lemma}[section]
\newtheorem{prop}{Proposition}[section]

\newtheorem*{thm*}{Theorem}

\usepackage{amsrefs}

\usepackage{hyperref, cleveref}

\title{Cohomological field theories and generalized Seiberg--Witten equations}
\author{Shuhan Jiang}
\author{J\"urgen Jost}
\affil[1]{Max Planck Institute for Mathematics in the Sciences, 04103 Leipzig}
\date{}

\begin{document}
	
	\maketitle
	
	\begin{abstract}
		We introduce a formalism for constructing cohomological field theories (CohFT) out of nonlinear PDEs based on the first author's previous work \cite{Jiang2023a}. We apply the formalism to the generalized Seiberg--Witten equations and show that the obtained CohFT functionals agree with the existing ones proposed by physicists. This leads to a unified perspective from which to view the full supersymmetric functionals of the Donaldson--Witten, Seiberg--Witten, and Kapustin--Witten theories and understand the relations between them. We also outline a quantization program for our framework and discuss its potential to produce manifold invariants and quantum cohomologies.
	\end{abstract}
	
    %\tableofcontents
	
	\section{Introduction}
	
	Gauge field theory has led to spectacular advances in geometry. It started with the groundbreaking work of Donaldson \cite{Donaldson83} who used moduli spaces of anti-selfdual connections as invariants for smooth structures on $4$-manifolds. See also \cite{Donaldson90} for a systematic exposition of the theory. The underlying gauge group here is non-abelian, for instance $\mathrm{SU}(2)$, and the gauge functional is the Yang--Mills functional. Later, the Seiberg--Witten equations were introduced \cites{Seiberg/Witten94a, Seiberg/Witten94b, Witten1994}:
	\begin{align}\label{SWeq}
		(F_A)_+ - \mu(\sigma)=0, \quad
		\slashed{D}_A  \sigma = 0.
	\end{align}
	Here $A$ is a connection on the determinant line bundle of a $\mathrm{spin}^c$ structure on a Riemannian $4$-manifold $M$, $\sigma$ is a section of the positive half-spinor bundle $S^+$ of the $\mathrm{spin}^c$-structure,  $\slashed{D}_A$ is the twisted Dirac operator, and $\mu: S^+ \rightarrow i \Lambda^2_+ T^*M$ is a fiber-wise quadratic bundle morphism. 
	%The gauge group here is the abelian group $\mathrm{U}(1)$, and 
	The resulting moduli spaces enabled a simpler approach than the original Donaldson theory. Still another such functional, the Kapustin--Witten functional \cite{Kapustin2007} relates its first-order field equations to the geometric Langlands program and  to non-abelian Hodge theory. For some recent development in the latter direction, see, for instance \cite{Liu2022}. Both Seiberg--Witten and Kapustin--Witten equations/functionals can be derived from the supersymmetric Yang--Mills theories and can be unified by the generalized Seiberg--Witten (GSW) equations/functionals \cites{Pidstrigach2004,Haydys2006,Ai2024}. The GSW equations are of the same form as \eqref{SWeq}, instead that one should interpret $A$ as a connection on a principal $G/\mathbb{Z}_2$-bundle associated with a $\mathrm{spin}^G$ structure on $M$, and $\sigma$ as a section of a generalized Clifford bundle (of positive chirality) of the $\mathrm{spin}^G$ structure.
	
	In this situation, it seems desirable to develop a unified perspective within which to situate these and possibly other gauge functionals. Naturally, this should be inspired by the underlying supersymmetric physics. 
	%Physicists' idea can be described as follows. 
	Let %$\mathcal{H} \hookrightarrow  \mathcal{E}_{tot} \rightarrow \mathcal{E}$
	\[
	\begin{tikzcd}
		\mathcal{H} \arrow[r, hook] & \mathcal{E}_{tot} \arrow[d,"\pi"] \\
		& \mathcal{E}
	\end{tikzcd}
	\]
	be a (trivial) $\mathcal{G}$-equivariant Fr\'echet vector bundle, where $\mathcal{E}:= \Gamma(E)$ is the space of sections of some fiber bundle $E \rightarrow M$, and $\mathcal{H}:=\Gamma(H)$ is the space of sections of some vector bundle $H \rightarrow M$. A (first-order) nonlinear PDE on $M$ can be often interpreted as a $\mathcal{G}$-equivariant section 
	\[
	\mathcal{F}: \mathcal{E} \rightarrow \mathcal{E}_{tot}.
	\]
	In the case of Seiberg--Witten equations, one has $\mathcal{E} = \mathcal{A} \times \Gamma(S^+)$, $\mathcal{H} = i \Omega^2_+(M) \times \Gamma(S^-)$, $\mathcal{E}_{tot} = \mathcal{E} \times \mathcal{H}$, and $\mathcal{G}$ is the gauge symmetry group. The $\mathcal{G}$-equivariant section $\mathcal{F}$ representing \eqref{SWeq} sends $(A, \sigma) \in \mathcal{E}$ to $((F_A)_+ - \mu(\sigma), \slashed{D}_A \sigma) \in \pi^{-1}((A,\sigma)) \cong \mathcal{H}$.
	%Let $\mathrm{Sol}({\mathcal{F}})$ denote the zero locus of $\mathcal{F}$, i.e., the solution space of the first-order nonlinear PDE represented by $\mathcal{F}$. The moduli space $\mathcal{M}({\mathcal{F}})$ of $\mathcal{F}$ is the quotient space $\mathrm{Sol}({\mathcal{F}})/\mathcal{G}$ and contains important information about the smooth structures of $M$.
	
	If $H$ is equipped with a $\mathcal{G}$-invariant bundle metric, one can define the following functional:
	\begin{align}\label{S}
		S := \int_M |\mathcal{F}|^2 \mathrm{vol}_M \in C^{\infty}(\mathcal{E}).
	\end{align}
	%whose critical locus $\mathrm{Crit}(S) \supset \mathrm{Sol}(\mathcal{F})$ can be described by a second-order nonlinear PDE. 
	\eqref{S} often admits a supersymmetric extension $\widetilde{S} \in C^{\infty}(\widetilde{\mathcal{E}})$ in the case of $M = \mathbb{R}^n$. Here $\widetilde{\mathcal{E}}$ is a Fr\'echet supermanifold with underlying manifold $\iota: \mathcal{E} \hookrightarrow \widetilde{\mathcal{E}}$. (In the case of Donaldson theory, \eqref{S} is nothing but the Yang--Mills functional up to a constant, and its supersymmetric extension $\widetilde{S}$ is the pure $N=2$ $\mathrm{SU}(2)$ supersymmetric Yang--Mills theory on $\mathbb{R}^4$.) Sometimes, $\widetilde{S}$ can be ``twisted" to be well-defined for a general Riemannian manifold $M$ at the cost of losing the full supersymmetries. In such case, 
	\begin{itemize}
		\item the supermanifold $\widetilde{\mathcal{E}}$ can be equipped with a compatible $\mathbb{Z}$-grading;
		\item $\widetilde{S}$ has degree $0$ and a remaining ``scalar'' supersymmetry $Q$ of degree $1$, $Q^2 = \frac{1}{2}[Q,Q]= 0$. 
	\end{itemize}
	Physicists refer to $(\widetilde{\mathcal{E}}, Q, \widetilde{S})$ as a cohomological field theory (CohFT) \cite{Witten1991}. The first example of a CohFT, the topologically twisted pure $N=2$ $\mathrm{SU}(2)$ supersymmetric Yang--Mills theory (also known as the Donaldson--Witten theory), was constructed by Witten to give a physical interpretation of Donaldson theory \cite{Witten1988}.
	
	It might appear, due to historical developments, that the cohomology of the complex $(C^{\infty}(\widetilde{\mathcal{E}}), Q)$ is strongly dependent on $\widetilde{S}$ and its supersymmetries. In this paper, however, we argue that this should not necessarily be the case. Based on the first author's previous work \cite{Jiang2023a}, we show that the complex $(C^{\infty}(\widetilde{\mathcal{E}}), Q)$ of a CohFT can be constructed solely from the data of its (first-order) field equations $\mathcal{F}$ and admits a clear mathematical interpretation. Moreover, we provide a universal expression for the supersymmetric action functional $\widetilde{S}$ in this setting. When applied to the GSW equations, our construction of $(\widetilde{\mathcal{E}}, Q, \widetilde{S})$ reproduces the full supersymmetric functionals of various CohFTs, such as the Donaldson--Witten, Seiberg--Witten, and Kapustin--Witten theories. Our formalism is closely related to, and can be seen as a refinement of the Mathai--Quillen formalism of CohFTs \cites{Mathai1986, Atiyah1990}. We refer the reader also to \cites{Blau1993,Wu2005} for wonderful reviews of the Mathai--Quillen formalism.	 
	
	The paper is organized as follows:
	
	In Section \ref{gc}, we provide a general procedure for constructing CohFTs based on the data of a nonlinear PDE represented by $\mathcal{F}$. We interpret/define $(C^{\infty}(\widetilde{\mathcal{E}}), Q)$ as the equivariant de Rham complex of the solution space of $\mathcal{F}$. Moreover, we establish a physics-mathematics dictionary for derived geometers, mathematical gauge theorists, and theoretical physicists.
    In Section \ref{0dtm}, we discuss a $0$-dimensional toy model and its relation to the Mathai--Quillen formalism and the Atiyah--Jeffrey formula. We also suggest a potential application of our framework to $n$-dimensional sigma model type theories for $n \geq 1$.
	In Section \ref{gsw}, we apply our construction to the generalized Seiberg--Witten equations in dimension $n=4$ and discuss the cases of Donaldson--Witten, Seiberg--Witten, and especially Kapustin--Witten theories in detail. 
	Section \ref{so} provides the mathematical definitions of symmetries and observables of a CohFT within our framework. In particular, we discuss two classes of symmetries of CohFTs in terms of our definition and compare them with the existing physics literature. We also show that the classical CohFT observables form a prefactorization algbera and discuss the traditional method of constructing CohFT observables using the so-called topological descent equations.
	Section \ref{qp} outlines a long-term $4$-step quantization program for our classical framework based on the existing perturbative quantization techniques developed by various mathematical physics communities centered around the Batalin--Vilkovisky (BV) formalism
	\cites{Cattaneo2014,Cattaneo2018,Costello2011,Costello2017,Costello2021,Rejzner2016}. 
	%The reason that BV formalism is applicable to our framework is that we are working with BRST model
	The final goal of this program is to provide a rigorous mathematical physics approach to defining manifold invariants and quantum cohomologies of a given CohFT. (The primary example that we have in mind is the Donaldson--Witten theory.) 
	Finally, Section \ref{cd} summarizes the main achievements of this paper and discusses other possible future research directions besides the ones listed in the quantization program of Section \ref{qp}.
	
	\section*{Acknowledgments}
	
	The first author would like to thank Owen Gwilliam for numerous discussions during his visit to UMass Amherst in May 2024, which not only helped crystallize his understanding of the constructions in his thesis on cohomological field theories but also inspired new constructions in the current paper. He would also like to thank Alberto Cattaneo, Chris Elliott, Andriy Haydys, Pavel Mnev, Alexander Schenkel, and Konstantin Wernli for their valuable communications, and thank the Max Planck Institute for Mathematics in the Sciences for their financial and administrative support.

	\section{The general construction}\label{gc}
	
	For every Lie algebra $\mathfrak{g}$, one can define a differential graded Lie superalgebra $\mathfrak{g}_{dR}  = \mathfrak{g}[-1] \oplus \mathfrak{g}$. The superbracket of $\mathfrak{g}_{dR}$ is induced by the bracket of $\mathfrak{g}$ and the adjoint action of $\mathfrak{g}$ on $\mathfrak{g}[-1]$. 
	%(The superbracket restricted to $\mathfrak{g}[-1]$ is 0.) 
	The differential $\delta_{dR}$ of $\mathfrak{g}_{dR}$ is given by
	\[
	0 \rightarrow \mathfrak{g}[-1] \xrightarrow{\mathrm{Id}}  \mathfrak{g} \rightarrow 0.
	\]
	Let $G$ be a Lie group with Lie algebra $\mathfrak{g}$. Let $(\mathcal{M}, Q_{\mathcal{M}})$ be a differential graded manifold equipped with a compatible $G$-action. (That is, for each $\xi \in \mathfrak{g}$, the fundamental vector field $X_{\xi}$ over $\mathcal{M}$ induced by $\xi$ commutes with $Q_{\mathcal{M}}$.) 
	%The graded commutative algebra $C^{\infty}(\mathcal{M})$ of functions over $\mathcal{M}$ is canonically a $\mathfrak{g}$-module. 
	%Let $\Omega(\mathcal{M})$ denote the space of differential forms on $\mathcal{M}$.
	The shifted tangent bundle $T[1]\mathcal{M}$ of $\mathcal{M}$, equipped with the differential $\delta_{\mathcal{M}}=d_{\mathcal{M}} + \mathrm{Lie}_{Q_{\mathcal{M}}}$, is a differential graded manifold. Here $d_{\mathcal{M}}$ is the de Rham differential of $\mathcal{M}$, and $\mathrm{Lie}_{Q_{\mathcal{M}}}$ is the Lie derivative of $Q_{\mathcal{M}}$. Let  $\mathfrak{X}(T[1]\mathcal{M})$ be the space of vector fields over $T[1]\mathcal{M}$. It is also a differential graded Lie superalgebra with the differential given by $[\delta_{\mathcal{M}}, \cdot]$.	
	One can check the map 
	\begin{align*}
		\rho: \mathfrak{g}_{dR} &\rightarrow \mathfrak{X}(T[1]\mathcal{M})\\
	    (\eta,\xi) &\mapsto  \iota_{X_{\eta}} + \mathrm{Lie}_{X_{\xi}},
	\end{align*}
	where $\iota_{X}$ is the contraction by a vector field $X$ over $\mathcal{M}$, defines a homomorphism between the differential graded Lie superalgebras $\mathfrak{g}_{dR}$ and $\mathfrak{X}(T[1]\mathcal{M})$. For example, one can check that
	\[
	[\rho((0,\xi)), \rho((\eta,0))] = [\mathrm{Lie}_{X_{\xi}}, \iota_{X_{\eta}}] = \iota_{[X_{\xi}, X_{\eta}]} = \iota_{X_{[\xi, \eta]}} = \rho([(0,\xi), (\eta,0)]),
	\]
	and
	\[
	[\delta_{\mathcal{M}}, \rho((\eta,0))] = [d_{\mathcal{M}} + \mathrm{Lie}_{Q_{\mathcal{M}}}, \iota_{X_{\eta}}] = \mathrm{Lie}_{X_{\eta}} + \iota_{[Q_{\mathcal{M}}, X_{\eta}]}= \mathrm{Lie}_{X_{\eta}}= \rho(\delta_{dR}((\eta,0))).
	\]
	%In other words, we have shown that
	\begin{lem}
		$\rho: \mathfrak{g}_{dR} \rightarrow \mathfrak{X}(T[1]\mathcal{M})$ defines a $\mathfrak{g}_{dR}$-module structure on $C^{\infty}(T[1]\mathcal{M})$.
	\end{lem}
	Recall that every differential graded Lie superalgebra can be viewed as a $L_{\infty}$ algebra with vanishing brackets $l_n$ for $n \geq 3$. Let $\mathcal{L}$ be an $L_{\infty}$ algebra and $(\mathcal{V}, \delta_{\mathcal{V}})$ be a (unital) differential graded commutative algebra. An $\mathcal{L}$-module structure on $\mathcal{V}$ is specified by a collection of graded skew-symmetric multilinear maps
	\[
	\rho_n: \mathcal{L}^{\otimes n-1} \otimes \mathcal{V} \rightarrow \mathcal{V}, \quad n \geq 1,
	\]
	of degree $2-n$ such that $\rho_1 = \delta_{\mathcal{V}}$ and
	\begin{align}
		\sum_{i+j=n} (-1)^{ij}  \sum_{\epsilon} (-1)^{\epsilon} \rho_{j+1}(\rho_i(x_{\epsilon(1)}, \dots, x_{\epsilon(i)}), x_{\epsilon(i+1)}, \dots, x_{\epsilon(n)}) = 0,\label{jacobi}
	\end{align}
	where $\epsilon$ ranges over $(i, j)$-unshuffles\footnote{Let $S(k)$ be the symmetric group of permutations of $\{1, \dots, k\}$. An element $\epsilon \in S(p+q)$ is an $(p,q)$-unshuffle if there exists $i_1 < \cdots <i_p$ and $j_1<\cdots<j_q$, such that $\epsilon(i_1)=1, \dots, \epsilon(i_p)=p$ and $\epsilon(j_1)=p+1, \dots, \epsilon(j_q)=p+q$.}, $i \geq 1$, $x_1, \dots, x_{n-1} \in \mathcal{L}$, and $x_{n} \in \mathcal{V}$. Note that either $\epsilon(n)$ or $\epsilon(i)$ equals $n$. If $\epsilon(n)=n$, one requires that $\rho_i(x_{\epsilon(1)}, \dots, x_{\epsilon(i)}) = l_i(x_{\epsilon(1)}, \dots, x_{\epsilon(i)})$.
	In particular, one can take $\mathcal{V}$ to be $\mathcal{L}$, $\delta_{\mathcal{V}}$ to be $l_1$, and all the higher brackets $\rho_n$ to be $l_n$.  \eqref{jacobi} then becomes the strong homotopy Jacobi identity of $\mathcal{L}$.

	The Chevalley--Eilenberg complex $(\mathrm{CE}(\mathcal{L}; \mathcal{V}), d_{CE})$ of $\mathcal{L}$ with values in $\mathcal{V}$ is defined as follows. %\cite[Appendix A]{Costello2017}.
	\begin{itemize}
		\item $\mathrm{CE}(\mathcal{L}; \mathcal{V}) = \mathrm{Sym}(\mathcal{L}^{\vee}[1]) \otimes \mathcal{V}$, where $\mathcal{L}^{\vee}$ is the dual graded vector space of $\mathcal{L}$;
		\item The differential $d_{CE}$ is determined by its restriction to the subspaces
		\[
		\mathrm{Sym}(\mathcal{L}^{\vee}[1]) \otimes \mathcal{V} \supset \mathrm{Sym}^1(\mathcal{L}^{\vee}[1]) \otimes 1 \cong \mathcal{L}^{\vee}[1], \quad \mathrm{Sym}(\mathcal{L}^{\vee}[1]) \otimes \mathcal{V} \supset 1 \otimes \mathcal{V} \cong \mathcal{V}.
		\]
		The Taylor coefficients 
		\[
		(d_{CE})_n: \mathcal{L}^{\vee}[1] \rightarrow \mathrm{Sym}^{n}(\mathcal{L}^{\vee}[1]) \otimes 1, \quad (d_{CE})_n: \mathcal{V} \rightarrow \mathrm{Sym}^{n-1}(\mathcal{L}^{\vee}[1]) \otimes \mathcal{V}
		\]
		are precisely the dual map of the higher bracket $l_n$ and the map $\rho_n$ regarded as an element in
		\[
		\mathrm{Hom}(\mathrm{Sym}^{n-1}(\mathcal{L}^{\vee}[1]), \mathrm{End}(\mathcal{V})) \cong \mathrm{Hom}(\mathcal{V}, \mathrm{Sym}^{n-1}(\mathcal{L}^{\vee}[1]) \otimes \mathcal{V}).
		\]
	\end{itemize}
	A straightforward computation shows that $d_{CE} \circ d_{CE}=0$ \cite[Section 4]{Reinhold2019}.
	
	\begin{rmk}
	In particular, $\mathfrak{g}_{dR}$ is an $L_{\infty}$ algebra and $C^{\infty}(T[1]\mathcal{M})$ is an $L_{\infty}$ module of $\mathfrak{g}_{dR}$. 
	\end{rmk}
	
	%10.8.2. Supersymmetrized Lie Algebra cohomology Greg Moore
	\begin{prop}\label{cewprop}
	    We have the following isomorphism of differential graded commutative algebras:
	    \begin{align}\label{cew}
	    	(\mathrm{CE}(\mathfrak{g}_{dR}; C^{\infty}(T[1]\mathcal{M})), d_{CE}) \cong (W(\mathfrak{g}) \otimes \Omega(\mathcal{M}), d_K).
	    \end{align}
	    Here $\Omega(\mathcal{M})$ is the graded commutative algebra of differential forms over $\mathcal{M}$, $W(\mathfrak{g}) = \Lambda(\mathfrak{g}^{\vee}) \otimes \mathrm{Sym}(\mathfrak{g}^{\vee})$ is the Weil algebra of $\mathfrak{g}$, and $d_K$ is the Kalkman differential
	    \[
	    d_K = d_W \otimes 1 + 1 \otimes \delta_{\mathcal{M}} + \theta^a \otimes \mathrm{Lie}_a - \phi^a \otimes \iota_a,
	    \]
	    where $\theta^a$ and $\phi^a$ are the degree $1$ and $2$ generators of $W(\mathfrak{g})$ induced by a basis $\{\xi_a\}$ of $\mathfrak{g}$, $\iota_a$ and $\mathrm{Lie}_a$ are the contractions and Lie derivatives on $\Omega(\mathcal{M})$ induced by $\xi_a$, and $d_W$ is the Weil differential
	    \[
	    d_W \theta^a = \phi^a - \frac{1}{2}f^a_{bc}\theta^b \theta^c, \quad d_W \phi^a = -f^a_{bc}\theta^b \phi^c,
	    \]
	    where $f^a_{bc}$ are the structure constants of $\mathfrak{g}$.
	\end{prop}
	\begin{rmk}
		If $Q_{\mathcal{M}}=0$, the right-hand side of \eqref{cew} is the BRST model of the equivariant de Rham cohomology of $\mathcal{M}$ \cites{Kalkman1993, Guillemin2013}.
	\end{rmk}
	\begin{proof}
		Note that 
		\[
		\mathrm{CE}(\mathfrak{g}_{dR}) = \mathrm{Sym}(\mathfrak{g}_{dR}^{\vee}[1]) = \mathrm{Sym}(\mathfrak{g}^{\vee}[(-1)(-1)+1] \oplus \mathfrak{g}^{\vee}[(-1)0+1]) \cong W(\mathfrak{g}),
		\] 
		and $C^{\infty}(T[1]\mathcal{M}) \cong \Omega(\mathcal{M})$. 
		Hence, $\mathrm{CE}(\mathfrak{g}_{dR}; C^{\infty}(T[1]\mathcal{M})) \cong W(\mathfrak{g}) \otimes \Omega(\mathcal{M})$. 
		It remains to show that $d_{CE} = d_K$ under this isomorphism. By definition, we have
		\[
		d_{CE} \theta^a = l_1^{\vee}(\theta^a) + l_2^{\vee}(\theta^a) = \phi^a - \frac{1}{2}f^a_{bc}\theta^b \theta^c, \quad d_{CE} \phi^a = l_1^{\vee}(\phi^a) + l_2^{\vee}(\phi^a) = 0 -f^a_{bc}\theta^b \phi^c,
		\]
		where $l_1 = \delta_{dR}$ and $l_2$ is the superbracket of $\mathfrak{g}_{dR}$. We also have
		\[
		d_{CE} \omega = \rho_1(\omega)  + \rho_2(\omega) = \delta_{\mathcal{M}} \omega - \phi^a \otimes \iota_a \omega + \theta^a \otimes \mathrm{Lie}_a \omega ,
		\]
		where $\rho_1$ and $\rho_2$ are determined by the $\mathfrak{g}_{dR}$-module structure $\rho$ of $\Omega(\mathcal{M})$. The minus sign of the above last term is due to the $-1$ degree of $\iota_a$. 
	\end{proof}
	
	Before applying this construction to study CohFTs, let us give a few more definitions. 
	
	Let $\mathcal{W}$ be a $\mathfrak{g}_{dR}$-algebra. 
	Let $\delta_{\mathcal{W}}$, $\iota_a$, and $\mathrm{Lie}_a$ denote the differential, contractions, and Lie derivatives on $\mathcal{W}$, respectively. An element $\Theta = \Theta^a \otimes \xi_a \in \mathcal{W} \otimes \mathfrak{g}$ of degree $1$ is called a connection of $\mathcal{W}$ if 
	\[
	\iota_a \Theta = \xi_a, \quad \mathrm{Lie}_a \Theta = -[\xi_a, \Theta].
	\]
	The curvature of $\Theta$ is an element $\Omega = \Omega^a \otimes \xi_a \in \mathcal{W} \otimes \mathfrak{g}$ of degree $2$ defined by the formula
	\[
	\Omega = \delta_{\mathcal{W}} \Theta + \frac{1}{2}[\Theta, \Theta].
	\]
	For example, the Weil algebra $W(\mathfrak{g})$ is a $\mathfrak{g}_{dR}$-algebra by setting
	\begin{align*}
		&\iota_a \theta^b = \delta_a^b, \quad \iota_a \phi^b = 0,\\
		&\mathrm{Lie}_a \theta^b = -f^b_{ac} \theta^c, \quad \mathrm{Lie}_a \phi^b = -f^b_{ac} \phi^c.
	\end{align*}
	The (unique) connection and curvature of the Weil algebra $W(\mathfrak{g})$ are given by the formulas $\theta = \theta^a \otimes \xi_a$ and $\phi = \phi^a \otimes \xi_a$. Let $P$ be a principal $G$-bundle. $\Omega(P)$ is also a $\mathfrak{g}_{dR}$-algebra, with $\iota_a$ and $\mathrm{Lie}_a$ given by the usual definitions of contractions and Lie derivatives. A connection/curvature of $\Omega(P)$ is simply a connection $1$-form/curvature $2$-form on $P$.
	%($\theta$ and $\phi$ are sometimes called the universal connection and curvature.) 	
	
	The Chern--Weil homomorphism 
	\[
    \mathrm{CW}_{\Theta}: W(\mathfrak{g}) \rightarrow \mathcal{W}
	\]
	is defined by sending
	\[
	\theta^a \mapsto \Theta^a, \quad \phi^a \mapsto \Omega^a.
	\]
	One can check that $\mathrm{CW}_{\Theta}$ is a morphism between $\mathfrak{g}_{dR}$-algebras. It sends the connection/curvature of $W(\mathfrak{g})$ to the connection/curvature of $\mathcal{W}$. For $\mathcal{W}=\Omega(P)$ and $\Theta$ a connection $1$-form on $P$, $CW_{\Theta}$ gives us the usual Chern--Weil homomorphism.

    %Elements in $\mathcal{W}$ that are $\mathfrak{g}_{dR}$-invariant are called basic elements. Elements in $\mathcal{W}$ that are only invariant under the action of the degree $-1$ part of $\mathfrak{g}_{dR}$ are called horizontal elemets.
		
	Let $\mathcal{W}$ and $\mathcal{W}'$ be two $\mathfrak{g}_{dR}$-algebras. Let $\Theta$ be a connection of $\mathcal{W}$. The Mathai--Quillen automorphism $T_{\Theta}$ of $\mathcal{W} \otimes \mathcal{W}'$ is defined as
	\[
	T_{\Theta} = \exp(\Theta^a \otimes \iota_a). 
	\]	
	Let $\mathcal{W} = W(\mathfrak{g})$ and $\mathcal{W}'= \Omega(\mathcal{M})$, where $\mathcal{M}$ is as in the Definition \ref{cewprop}. Let $\Theta = \theta$ be the canonical connection of $W(\mathfrak{g})$. One has \cite{Kalkman1993}
	\begin{align*}
		&T_{\theta} \circ d_K \circ T_{\theta}^{-1} = d_W \otimes 1  + 1 \otimes \delta_{\mathcal{M}}, \\
		&T_{\theta} \circ \iota_a \otimes 1 \circ T_{\theta}^{-1} = \iota_a \otimes 1 + 1 \otimes \iota_a, \\
		&T_{\theta} \circ (\mathrm{Lie}_a \otimes 1 + 1 \otimes \mathrm{Lie}_a) \circ T_{\theta}^{-1} =\mathrm{Lie}_a \otimes 1 + 1 \otimes \mathrm{Lie}_a.
	\end{align*}
	Therefore, $\mathrm{CE}(\mathfrak{g}_{dR}; C^{\infty}(T[1]\mathcal{M}))$ can be equipped with an $\mathfrak{g}_{dR}$-algebra structure via the isomorphism \eqref{cew}.

	\subsection{The minimal construction of CohFTs}\label{minCohFT}
	
	To apply the above construction to study CohFTs, we consider the following Koszul complex
	\[
	\cdots \xrightarrow{\iota_{\mathcal{F}}} \Gamma(\Lambda^{k} \mathcal{E}_{tot}^{\vee}) \xrightarrow{\iota_{\mathcal{F}}} \cdots  \xrightarrow{\iota_{\mathcal{F}}}  \Gamma(\mathcal{E}_{tot}^{\vee}) \xrightarrow{\iota_{\mathcal{F}}} C^{\infty}(\mathcal{E}) \rightarrow 0,
	\]
	where $\mathcal{E}_{tot}^{\vee}$ is the dual bundle\footnote{The fiber $\mathcal{H}^{\vee}$ of $\mathcal{E}_{tot}^{\vee}$ is the continuous liner dual of the fiber $\mathcal{H}$ of $\mathcal{E}_{tot}$.} of the Fr\'echet vector bundle $\mathcal{E}_{tot}$ over $\mathcal{E}$, and $\iota_{\mathcal{F}}$ is the contraction by the $\mathcal{G}$-equivariant section $\mathcal{F}: \mathcal{E} \rightarrow \mathcal{E}_{tot}$. This complex is equivalent to the infinite dimensional differential graded manifold $(\mathcal{E}_{tot}[-1], \iota_{\mathcal{F}})$. $\mathcal{E}_{tot}[-1]$ is equipped with a $\mathcal{G}$-action, which is compatible with its cohomological vector field $\iota_{\mathcal{F}}$ because $\mathcal{F}$ is $\mathcal{G}$-equivariant. 
	\begin{defn}
		The minimal CohFT extension $(\widetilde{\mathcal{E}}_{min}, Q, \widetilde{S}_{min})$ of $(\mathcal{E}, S)$ is given by
		\begin{itemize}
			\item $\widetilde{\mathcal{E}}_{min} = T[1] (\mathrm{Lie}(\mathcal{G})[1]) \times T[1](\mathcal{E}_{tot}[-1])$;
			\item $Q$ is the Chevalley--Eilenberg differential under the isomorphism 
			\begin{align}\label{miniso}
				C^{\infty}(\widetilde{\mathcal{E}}_{min}) \cong \mathrm{CE}(\mathrm{Lie}(\mathcal{G})_{dR}; C^{\infty}(T[1](\mathcal{E}_{tot}[-1])));
			\end{align}
			\item Let $(\Phi, \chi)$ of degree $(0,-1)$ denote local coordinates of $\mathcal{E}_{tot}[-1]$ and $(\Psi, b)$ of degree $(1, 0)$ denote the local coordinates of its shifted tangent space. The minimal CohFT action functional $\widetilde{S}_{min}$ is defined as
			\[
			\widetilde{S}_{min} = Q \left(\int_M \langle \chi, b \rangle \mathrm{vol}_M \right).
			\]
		\end{itemize}
	\end{defn}
	
	\begin{rmk}
		The underlying manifold of $\widetilde{\mathcal{E}}_{min}$ is not $\mathcal{E}$, but $\mathcal{E}_{tot}$. The underlying even manifold of $\widetilde{\mathcal{E}}_{min}$ is $\mathrm{Lie}(\mathcal{G})[2] \times \mathcal{E}_{tot}$. We use $\iota$ and $\iota_{even}$ to denote their embeddings into $\widetilde{\mathcal{E}}_{min}$, respectively.
	\end{rmk}
	
	\begin{prop}
		The pullback 
		\begin{align}\label{fofS}
			\iota^* \widetilde{S}_{min} = \int_M \left(|b|^2 + \langle \mathcal{F}, b \rangle \right) \mathrm{vol}_M =: S_{fo}
		\end{align}
		provides a first-order formulation of $S$.
	\end{prop}
	It is easy to see that $\mathrm{Crit}(S_{fo}) \cong \mathrm{Crit}(S)$ and that
		\begin{align}\label{fofSsol}
			\mathrm{Crit}(S_{fo}) \cap S_{fo}^{-1}(0) \cong \mathrm{Sol}(\mathcal{F}),
		\end{align}
	where we use $\mathrm{Sol}({\mathcal{F}})$ denote the solution space of the PDEs represented by $\mathcal{F}$. 
 	\begin{proof}
		Let $(\theta, \phi)$ denote the coordinates of $T[1](\mathrm{Lie}(\mathcal{G})[1]) \cong \mathrm{Lie}(\mathcal{G})[1] \oplus \mathrm{Lie}(\mathcal{G})[2]$. The scalar supersymmetry $Q$ is defined by its action on the fields:
		\begin{align*}
			&Q \theta = \phi - \frac{1}{2}[\theta,\theta],~Q\phi=-[\theta,\phi],\\
			&Q \Phi = \Psi - \theta \Phi, ~ Q \Psi = -\theta \Psi + \phi \Phi,\\
			&Q \chi = b - \theta \chi + \mathcal{F}(\Phi), ~ Qb = -\theta b + \phi \chi - \mathrm{Lin}_{\Phi}(\mathcal{F}) \Psi,
		\end{align*}
		where we use $\mathrm{Lin}_{\Phi}(\mathcal{F})$ to denote the linearization of $\mathcal{F}$ at $\Phi$ and $\theta\Phi$ to denote the action of $\theta$ on $\Phi$.
		A direct computation shows that
		\[
		\widetilde{S}_{min} = \int_M \left(|b|^2 + \langle \mathcal{F}, b \rangle + \langle \chi, \mathrm{Lin}_{\Phi}(\mathcal{F})\Psi \rangle - \langle \chi, \phi \chi \rangle \right) \mathrm{vol}_M.
		\]
		%This completes the proof.
	\end{proof}
		
	\begin{prop}
		$\widetilde{S}_{min}$ is $\mathrm{Lie}(\mathcal{G})_{dR}$-invariant.
	\end{prop}
	\begin{proof}
		This follows directly from the fact that $\int_M \langle \chi, b \rangle \mathrm{vol}_M$ is $\mathcal{G}$-invariant and independent of $\theta$.
	\end{proof}
	
	For a derived geometer, the differential graded manifold $(\mathcal{E}_{tot}[-1], \iota_{\mathcal{F}})$ provides a concrete model for the derived manifold $\mathrm{Sol}(\mathcal{F})$ of the intersection of $\mathcal{F}$ and the zero section \cites{Carchedi2019,Carchedi2023}. From this point of view, the study of a minimal CohFT $(\widetilde{\mathcal{E}}_{min}, Q, \widetilde{S}_{min})$ is equivalent to the study of the $\infty$-dimensional equivariant ``de Rham" cohomology theory of $\mathrm{Sol}(\mathcal{F})$.\footnote{$\widetilde{S}_{min}$ plays no role since its first-order bosonic field equations coincide with the first-order PDE $\mathcal{F}$. See \eqref{fofSsol}.} We use quotation marks for ``de Rham" because $\mathrm{Sol}(\mathcal{F})$ is not necessarily smooth. However, if we are lucky enough that 
	\begin{enumerate}
		\item $\mathcal{F}$ is transversal; 
		\item the action of $\mathcal{G}$ on $\mathrm{Sol}(\mathcal{F})$ is free and admits local slices;
		\item letting 
		\[
		0 \rightarrow \mathrm{Lie}(\mathcal{G}) \xrightarrow{R_{\Phi}} T_{\Phi} \mathcal{E} \xrightarrow{\mathrm{Lin}_{\Phi}(\mathcal{F})} VT_{(\Phi,0)} \mathcal{E}_{tot} \cong \mathcal{H} \rightarrow 0
		\]
		be the deformation complex of $\mathcal{F}$ at $\Phi \in \mathrm{Sol}{\mathcal{F}}$, where $R_{\Phi}$ is the linearization of the right action of $\mathcal{G}$ at $\Phi$ and $VT \mathcal{E}_{tot}$ is the vertical bundle over $\mathcal{E}_{tot}$, the map
		\[
		\mathcal{D}_{\Phi}:=(R_{\Phi}^*, \mathrm{Lin}_{\Phi}(\mathcal{F})): T_{\Phi} \mathcal{E} \rightarrow \mathrm{Lie}(\mathcal{G}) \oplus \mathcal{H}
		\]
		is Fredholm, where $R_{\Phi}^*$ is the formal adjoint of $R_{\Phi}$;
	\end{enumerate}
	then the moduli space $\mathcal{M}(\mathcal{F}) = \mathrm{Sol}(\mathcal{F})/\mathcal{G}$ is a smooth manifold of finite dimension $\mathrm{Ind}(\mathcal{D}_{\Phi})$. (This is the standard approach taken by the mathematical gauge theory community. See, for example, \cite[Theorem 183.]{Haydys2019}.) In other words, we may say that: 
	
	\vspace{5pt}
	In good cases, the study of the minimal CohFT $(\widetilde{\mathcal{E}}_{min}, Q, \widetilde{S}_{min})$ constructed out of $\mathcal{F}$ is the study of the de Rham cohomology of the moduli space $\mathcal{M}(\mathcal{F})$.
    \vspace{5pt}
    
    For a theoretical physicist, she or he may interpret our construction of a CohFT as the ``de Rham version'' of a BRST theory. The degrees of the fields  in our construction can be interpreted as the ghost number degrees. Recall that the BRST complex is the Chevalley--Eilenberg complex $\mathrm{CE}(\mathrm{Lie}(\mathcal{G}); C^{\infty}(\mathcal{E}_{tot}[-1]))$.\footnote{In our construction of minimal CohFTs, we replace $\mathrm{Lie}(\mathcal{G})$ with $\mathrm{Lie}(\mathcal{G})_{dR}$ and $\mathcal{E}_{tot}[-1]$ with $T[1](\mathcal{E}_{tot}[-1])$. }  In this setting, one only has three fields: $\theta$, $\Phi$, and $\chi$. And the BRST operator $Q_{BRST}$ is given by
    \[
    Q_{BRST} \theta = -\frac{1}{2}[\theta, \theta], ~ Q_{BRST} \Phi = \theta \Phi, ~ Q_{BRST} \chi = \mathcal{F}(\Phi) + \theta \chi.
    \]
    The BRST action functional of $\mathcal{F}$ is defined as
    \[
    S_{BRST} = Q_{BRST}\left(\int_M \langle \chi, \mathcal{F}(\Phi) \rangle \mathrm{vol}_M \right).
    \]
    It is easy to show that
    \[
    S_{BRST} = \int_M |\mathcal{F}|^2 \mathrm{vol}_M = S,
    \]
    which is not the correct BRST action functional in the physics literature. This is because we are implicitly working with the zero function $0$ (or more generally, a component-wise constant function) on $\mathcal{E}$ and using $\mathcal{F}$ to gauge fix its symmetry group, the (identity component of the) diffeomorphism group of $\mathcal{E}$ \cite{Baulieu1988}. While in the standard BRST formalism, one starts with $S=\int_M |\mathcal{F}|^2 \mathrm{vol}_M$ and uses a section
    \[
    gf: \mathcal{E} \rightarrow \mathcal{E} \times \mathcal{H}'
    \]
    to gauge fix the $\mathcal{G}$-symmetry of $\mathrm{Crit}(S)$. For example, if $\mathcal{E}$ is the space of $G$ connections $A$ on $M = \mathbb{R}^n$, one can chose $\mathcal{H}' = \mathrm{Lie}(\mathcal{G}) = C^{\infty}(M,\mathfrak{g})$ and the Coulomb gauge fixing:
    \[
    gf(A) = d^*A,
    \]
    where $d^*$ is the formal adjoint of the exterior derivative $d$ of $M$.
    %, to gauge fix the $C^{\infty}(M,G)$-symmetry of $\mathrm{Crit}(S)$. 
    %Since $gf$ is not $\mathcal{G}$-equivariant, one needs to introduce two fields: $c$ and $\overline{c}$ of degrees $-1$ and $0$.
   
    %When applying to the anti-self dual equations, the ordinary BRST construction will produce the self-dual Yang--Mills theory, while our construction will produce a minimal version of the Donaldson--Witten theory.

    \subsection{The standard construction of CohFTs}
    
    Historically, CohFTs were first constructed by physicists by applying topological twists to supersymmetric field theories. Our minimal construction in the previous subsection, though quite neat, does not fully recover some of the physicist's CohFTs. In this subsection, we fill in this gap by adding an additional sector to the minimal construction. The inclusion of this additional sector can tell us more information about the part of $\mathrm{Sol}(\mathcal{F})$ where $\mathcal{G}$ acts non-freely.
	
    \begin{defn}
    	The (standard) CohFT extension $(\widetilde{\mathcal{E}}, Q, \widetilde{S})$ of $(\mathcal{E}, S)$ is given by
    	\begin{itemize}
    		\item $\widetilde{\mathcal{E}} = T[1] (\mathrm{Lie}(\mathcal{G})[1]) \times T[1](\mathcal{E}_{tot}[-1] \times \mathrm{Lie}(\mathcal{G})[-2])$;
    		%where we view $\mathrm{Lie}(\mathcal{G})[-2] $ as a trivial Fr\'echet vector bundle over $\mathcal{E}$ and $\oplus_{\mathcal{E}}$ is the Whitney sum;
    		\item $Q$ is the Chevalley--Eilenberg differential under the isomorphism
    		\begin{align}\label{isostand}
    			C^{\infty}(\widetilde{\mathcal{E}} ) \cong \mathrm{CE}(\mathrm{Lie}(\mathcal{G})_{dR}; C^{\infty}(T[1](\mathcal{E}_{tot}[-1] \times \mathrm{Lie}(\mathcal{G})[-2]))),
    		\end{align}
    		where we equip $T[1](\mathcal{E}_{tot}[-1] \times \mathrm{Lie}(\mathcal{G})[-2]))$ with the cohomological vector field $\delta_{\mathcal{E}_{tot}[-1]} + d_{\mathrm{Lie}(\mathcal{G})[-2]}$, $d_{\mathrm{Lie}(\mathcal{G})[-2]}$ is the de Rham differential of $\mathrm{Lie}(\mathcal{G})[-2]$;
    		\item Let $(\lambda, \eta)$ of degree $(-2,-1)$ denote the coordinates of $T[1](\mathrm{Lie}(\mathcal{G})[-2]) \cong \mathrm{Lie}(\mathcal{G})[-2] \oplus \mathrm{Lie}(\mathcal{G})[-1]$. The (standard) CohFT extension $\widetilde{S}$ of $S$ is given by
    		\[
    		\widetilde{S} = Q \left(\int_M \left(\langle \chi, b \rangle + \langle \Psi, \lambda \Phi \rangle + s \langle \eta, \mathrm{ad}_{\phi}(\lambda)\rangle \right)\mathrm{vol}_M \right),
    		\]
    		where we use $\lambda \Phi$ to denote the action of $\lambda$ on $\Phi$, use $\mathrm{ad}_{\phi}(\lambda)$ to denote the adjoint action of $\phi$ on $\lambda$, $s$ is a non-zero real parameter. (If not otherwise explicitly stated, $s$ shall be set to be $1$.)
    	\end{itemize}
    \end{defn}
    
    \begin{rmk}
    	The underlying manifold of $\widetilde{\mathcal{E}}$ is still $\mathcal{E}_{tot}$, same as the underlying manifold of $\widetilde{\mathcal{E}}_{min}$. The underlying even manifold of $\widetilde{\mathcal{E}}$ is $\mathrm{Lie}(\mathcal{G})[2] \times \mathcal{E}_{tot} \times \mathrm{Lie}(\mathcal{G})[-2]$.  Since the newly introduced fields $\eta$ and $\lambda$ have nonzero degrees, $\iota^* \widetilde{S}$ still takes the form \eqref{fofS}, but $\widetilde{S}_{even}:=\iota_{even}^* \widetilde{S}$ has now more terms.
    \end{rmk}
    
    \begin{prop}
    	The bosonic part $\widetilde{S}_{even}$ of $\widetilde{S}$ can be written as
    	\begin{align}\label{Seven}
    		\widetilde{S}_{even}= S_{fo} + \int_M \left(\langle \phi \Phi, \lambda \Phi \rangle + |[\phi, \lambda]|^2 \right) \mathrm{vol}_M.
    	\end{align}
    \end{prop}
    
    \begin{proof}
    	The actions of $Q$ on $\lambda$ and $\eta$ are given by
    	\[
    	Q \lambda = \eta - [\theta, \lambda],~ Q \eta = - [\theta, \eta] + [\phi, \lambda].
    	\]
    	Direct computations show that
    	\[
    	\widetilde{S} = \widetilde{S}_{min} + \int_M \left(\langle \phi \Phi, \lambda \Phi \rangle + |[\phi, \lambda]|^2 \right) \mathrm{vol}_M -  \int_M \left(\langle \Psi, \eta \Phi \rangle + \langle \Psi, \lambda \Psi \rangle  + \langle \eta,  [\phi, \eta] \rangle \right) \mathrm{vol}_M.
    	\]
    	This completes the proof.
    \end{proof}
    
    Likewise, one can show that
    \begin{prop}
    	$\widetilde{S}$ is $\mathrm{Lie}(\mathcal{G})_{dR}$-invariant.
    \end{prop}
    
    The first-order field equations of $\widetilde{S}_{even}$ are given by
    \[
    \mathcal{F}(\Phi) = 0, \quad \lambda \Phi = \phi \Phi = 0, \quad [\phi, \lambda]=0.
    \]
    They describe the critical locus of $\widetilde{S}_{even}$ of value $0$:
    \[
    \widetilde{\mathrm{Sol}(\mathcal{F})}:=\mathrm{Crit}(\widetilde{S}_{even}) \cap \widetilde{S}_{even}^{-1}(0).
    \]
    Note that there exists a canonical projection
    \[
    \mathrm{pr}: \widetilde{\mathrm{Sol}(\mathcal{F})} \rightarrow \mathrm{Sol}(\mathcal{F})
    \]
    defined by sending $\phi$ and $\lambda$ to $0$.
    Let $\mathrm{Sol}(\mathcal{F})_{irred}$ denote the subspace of $\mathrm{Sol}(\mathcal{F})$ on which $\mathcal{G}$ acts freely, and $\mathrm{Sol}(\mathcal{F})_{red}$ denote the subspace of $\mathrm{Sol}(\mathcal{F})$ on which $\mathcal{G}$ acts non-freely. The fiber
    \[
    \mathcal{R}_{\Phi}:=\mathrm{pr}^{-1}(\Phi)=\{(\phi, \lambda)| \phi \Phi = \lambda \Phi = 0, [\phi, \lambda]=0\}
    \]
    of the projection has a canonical complex structure defined by
    \[
    I(\phi) = -\lambda, \quad I(\lambda) = \phi.
    \]
    $\mathcal{R}_{\Phi}$ is nontrivial if and only if $\Phi \in \mathrm{Sol}(\mathcal{F})_{red}$.
    %We can draw the following picture
    
    \begin{figure}[h]
    	\centering
    	\begin{tikzpicture}   		
    		% Label the manifold
    		\node at (1, -0.5) {$\mathrm{Sol}(\mathcal{F})_{red}$};
    		
    		% Draw a point on the manifold
    		\fill[black] (-2, 0.5) circle (2pt) node[below] {$(\Phi,0)$};
    		
    		% Draw a fiber (vector space) over the point
    		\fill[gray!20] (-2, 0.5) -- (-2, 2.5) -- (0, 3.5) -- (0, 1.5) -- cycle; % Fill the parallelogram with gray and some opacity
    		
    		% Define a smooth, wavy shape for the manifold
    		\draw [thick, smooth] plot coordinates {
    			(-4, 0) (-3.5, 0.5) (-3, 1) (-2.5, 1.2) (-2, 1.4) (-1.5, 1.5) (-1, 1.4)
    			(-0.5, 1.2) (0, 1) (0.5, 0.8) (1, 0.6) (1.5, 0.4) (2, 0.2) (2.5, 0)
    			(3, -0.2) (3.5, -0.4) (4, -0.5)
    		};
    		
    		% Draw the mirrored wavy shape to form a closed loop
    		\draw [thick, smooth] plot coordinates {
    			(4, -0.5) (3.5, -1) (3, -1.4) (2.5, -1.7) (2, -1.8) (1.5, -1.9) 
    			(1, -1.8) (0.5, -1.7) (0, -1.6) (-0.5, -1.4) (-1, -1.2) (-1.5, -1) 
    			(-2, -0.8) (-2.5, -0.6) (-3, -0.4) (-3.5, -0.2) (-4, 0)
    		};
    		
    		\draw[thick, ->] (-2, 0.5) -- (-2, 2.5) node[above] {$\lambda$};
    		\draw[thick, ->] (-2, 0.5) -- (0, 1.5) node[right] {$\phi$};
    		
    		% Mark the fiber
    		\node at (-1, 2) {$\mathcal{R}_{\Phi}$};  % Label for the fiber
    	\end{tikzpicture}
    	\caption{A picture of $\mathrm{pr}^{-1}(\mathrm{Sol}(\mathcal{F})_{red}) \subset \widetilde{\mathrm{Sol}(\mathcal{F})}$.}
    	\label{fig:pic1}
    \end{figure}
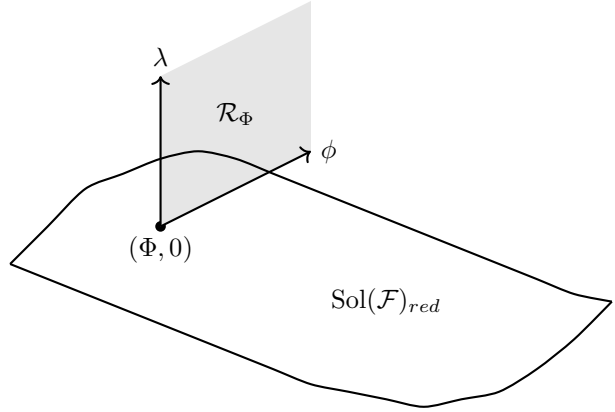
    
    \begin{exmp}
    	If $M=\mathbb{R}^4$ and $\mathcal{F}=(F_A)_+$, the quotient space $\mathcal{R}_{A=0}/\mathcal{G}$ is isomorphic to $\mathbb{C}$ and is known as the classical $u$-plane of the Donaldson--Witten theory \cite{Moore1997}. %The Coulomb branch of the theory is isomorphic to $\mathbb{C}-\{0\}$.
    \end{exmp}
    
    Sometimes, one has the following decomposition:
    \[
    \mathcal{E}_{tot} = \mathcal{E}_{tot,1} \times \mathcal{E}_{tot,2}
    \]
    such that the base manifold $\mathcal{E}_2$ of $\mathcal{E}_{tot,2}$ is a Fréchet vector space on which $\mathcal{G}$ acts freely everywhere except at $0 \in \mathcal{E}_2$. 
    Let's consider the PDE represented by the following $\mathcal{G}$-equivariant section
    \[
    \mathcal{F}_{\mu}(\Phi_1, \Phi_2) = (\mathcal{F}_1(\Phi_1) - \frac{1}{2}\mu(\Phi_2), \mathcal{F}_2(\Phi_2)) \quad (\Phi_1, \Phi_2) \in \mathcal{E}_1 \times \mathcal{E}_2,
    \]
    where $\mathcal{F}_1$ and $\mathcal{F}_2$ contain all the derivative terms of $\mathcal{F}_{\mu}$, and $\mu$ is a potential term satisfying $\mu(0)=0$. ($\mu$ typically quadratic in $\Phi_2$.) Then the solution space $\mathrm{Sol}(\mathcal{F}_{\mu})$ can be decomposed as
    \[
    \mathrm{Sol}(\mathcal{F}_{\mu}) = \mathrm{Sol}(\mathcal{F}_{\mu})_{\mu=0} \sqcup \mathrm{Sol}(\mathcal{F}_{\mu})_{\mu \neq 0} =  \mathrm{Sol}(\mathcal{F}_1) \times \left(\mathrm{Sol}(\mathcal{F}_2) \cap \{\Phi_2 \in \mathcal{E}_2| \mu(\Phi_2)=0\} \right) \sqcup \mathrm{Sol}(\mathcal{F}_{\mu})_{\mu \neq 0}.
    \]
    Obviously, we have 
    \[
    \mathrm{Sol}(\mathcal{F}_{\mu})_{red} \subset \mathrm{Sol}(\mathcal{F}_{\mu})_{\mu=0}.
    \]
    %and $\mathrm{Sol}(\mathcal{F}_{\mu})_{red}  \cong \mathrm{Sol}(\mathcal{F})_{red}$. 
    With a slight abuse of notation, we use $\mathrm{pr}$ to denote the composition of the projections
    \[
    \widetilde{\mathrm{Sol}(\mathcal{F}_{\mu})_{\mu=0}} \rightarrow \mathrm{Sol}(\mathcal{F}_{\mu})_{\mu=0} \rightarrow \mathrm{Sol}(\mathcal{F}_1).
    \]
    For $\Phi_1 \in \mathrm{Sol}(\mathcal{F}_{1})$, we denote
    \[
    \mathcal{R}_{\Phi_1} := \mathrm{pr}^{-1}(\Phi_1).
    \]
    The freeness of the $\mathcal{G}$-action on $\mathcal{E}_2$ and the equations $\phi \Phi_2 = \lambda \Phi_2 = 0$ imply that
    \[
    \mathcal{R}_{\Phi_1}  = \mathcal{R}_{\Phi_1}^{\phi=\lambda=0} \oplus \mathcal{R}_{\Phi_1}^{\Phi_2=0},
    \]
    where
    \[
     \mathcal{R}_{\Phi_1}^{\phi=\lambda=0}  = \{\Phi_2| \mathcal{F}_2(\Phi_2)=0, \mu(\Phi_2)=0\}, \quad  \mathcal{R}_{\Phi_1}^{\Phi_2=0} = \{(\phi, \lambda)| \phi \Phi_1 = \lambda \Phi_1 = 0, [\phi, \lambda]=0\}.
    \]
    (Clearly, $\mathcal{R}_{\Phi_1}^{\Phi_2=0}$ is nontrivial if and only if $\Phi_1 \in \mathrm{Sol}(\mathcal{F}_1)_{red}$.)
    In the cases where $M=\mathbb{R}^n$ and $\mathcal{E}_1$ is also a Fr\'echet vector space, physicists refer to the quotient $\mathcal{R}_{\Phi_1=0}/\mathcal{G}$ as the moduli spaces of classical vacua, and refer to the quotients
    \[
    \mathcal{R}_{\Phi_1=0}^{\phi=\lambda=0}/\mathcal{G}~ \mathrm{and}~ \mathcal{R}_{\Phi_1=0}^{\Phi_2=0}/\mathcal{G}
    \]
     as the classical Higgs branch and the classical Coulomb branch of the CohFT, respectively \cite{Moore2017}.
     \begin{exmp}
     	The Coulomb branch of the Donaldson--Witten theory is isomorphic to $\mathbb{C}-\{0\}$.
     	%and Seiberg--Witten
     \end{exmp}
     
     %\subsection{The extended construction of CohFTs}
     
     %\subsubsection{The standard construction as an extended minimal construction}
     
     %$\mathbb{Z} \times \mathbb{Z}$-grading.

     %\subsection{The 0-dimensional toy model, localization, and Poincar\'e--Hopf theorem}
     %\section{The 0-dimensional toy model and Mathai--Quillen formalism}
     
     \section{The 0-dimensional toy model}\label{0dtm}
     
     In this section, we consider the zero dimensional toy model where $M$ is a point, $\mathcal{E}$ is the frame bundle $\mathrm{Fr}(N)$ over an $2m$-dimensional Riemannian manifold $N$, $\mathcal{H}=\mathbb{R}^{2m}$, $\mathcal{E}_{tot}$ is $\mathrm{Fr}(N) \times \mathbb{R}^{2m}$, and $\mathcal{G} = \mathrm{SO}(2m)$. Note that the $\mathcal{G}$-action on $\mathcal{E}_{tot}$ is free and we have 
     \[
     \mathcal{E}_{tot}/\mathcal{G} \cong TN.
     \]
     Therefore, an $\mathcal{G}$-equivariant section $\mathcal{F}$ of $\mathcal{E}_{tot}$ is equivalent to a vector field over $N$. 
     
     \subsection{Mathai--Quillen formalism}
         
     The configuration space of the $0$-dimensional minimal CohFT is 
     \[
     \widetilde{\mathcal{E}}_{min} = T[1](\mathfrak{so}(2m)[1]) \times T[1](\mathrm{Fr}(N) \times \mathbb{R}^{2m}[-1]).
     \]
     %We have
     %$
     %C^{\infty}(\widetilde{\mathcal{E}}_{\min}) \cong W(\mathfrak{so}(2m)) \otimes \Omega(\mathrm{Fr}(N)) \otimes \Omega (\mathbb{R}^{2m}[-1]).
     %$
     Let 
     $\Delta: \mathrm{Fr}(N) \rightarrow \mathrm{Fr}(N) \times \mathrm{Fr}(N)$
     denote the diagonal embedding. Let 
     $\mathrm{CW}_{\nabla}: W(\mathfrak{so}(2m)) \rightarrow \Omega(\mathrm{Fr}(N))$ 
     denote the Chern--Weil homomorphism induced by the Levi-Civita connection $\nabla$ of $N$. Let $T_{\nabla}$ denote the Mathai--Quillen automorphism of $\Omega(\mathrm{Fr}(N)) \otimes \Omega(\mathrm{Fr}(N))$ induced by $\nabla$. $\Delta$, $\mathrm{CW}_{\nabla}$, and $T_{\nabla}$ together induce a homomorphism $J$ between $\mathfrak{so}(2m)_{dR}$-algebras:
     \begin{align}
     	J: &C^{\infty}(\widetilde{\mathcal{E}}_{\min}) \cong W(\mathfrak{so}(2m)) \otimes \Omega(\mathrm{Fr}(N)) \otimes \Omega(\mathbb{R}^{2m}[-1]) \xrightarrow{(CW_{\nabla} \otimes 1 \otimes 1)}  \notag \\
     	&\left( \Omega(\mathrm{Fr}(N)) \otimes \Omega(\mathrm{Fr}(N)) \right) \otimes \Omega(\mathbb{R}^{2m}[-1]) \xrightarrow{T_{\nabla} \otimes 1} \notag \\
     	&\left( \Omega(\mathrm{Fr}(N)) \otimes \Omega(\mathrm{Fr}(N)) \right) \otimes \Omega(\mathbb{R}^{2m}[-1]) \xrightarrow{\Delta^* \otimes 1} \notag \\
     	&\Omega(\mathrm{Fr}(N)) \otimes \Omega(\mathbb{R}^{2m}[-1])
     	\cong C^{\infty}(\widetilde{\mathcal{E}}_{MQ}), \label{J}
     \end{align}
     where 
     \[
     \widetilde{\mathcal{E}}_{MQ}:= T[1](\mathrm{Fr}(N) \times \mathbb{R}^{2m}[-1]).
     \]
     The cohomological vector field $Q_J := J \circ Q \circ J^{-1}$ of $\widetilde{\mathcal{E}}_{MQ}$ is given by
     \begin{align}\label{nongau}
     	 Q_J \Phi = \Psi, \quad Q_J \Psi = 0, \quad Q_J \chi = b + \mathcal{F} - A_{\nabla} \chi, \quad Q_J b = -A_{\nabla} b + R_{\nabla} \chi - \nabla \mathcal{F}, 
     \end{align}
     where $A_{\nabla}$ is the connection $1$-form of $\nabla$, and $R_{\nabla}=d A_{\nabla} + \frac{1}{2}[A_{\nabla}, A_{\nabla}]$ is the Riemannian curvature of $N$.
     %, viewed as a $\mathrm{so}(2m)$-valued $2$-form on $\mathrm{Fr}(N)$.
     % One can also check that its contractions and Lie derivatives are given by the standard definitions.
     
     The image of the minimal CohFT action functional $\widetilde{S}_{min}$ under $J$ is given by
     \[
     \widetilde{S}_{MQ}:=J(\widetilde{S}_{min}) = Q_J J (\langle \chi, b \rangle) = Q_J \langle \chi, b \rangle =  |b|^2 + \langle \mathcal{F}, b \rangle + \langle \chi, \nabla \mathcal{F} \rangle - \langle \chi, R_{\nabla} \chi \rangle.
     \]
     %where and $\langle \cdot, \cdot \rangle$ is the Riemannian metric of $N$.
     For the reader's convenience, we recall the following formulas of the Gaussian type Berezin integrals:
     \begin{align*}
     	&\int d b \exp(-\frac{1}{2}\langle b, \mathcal{A} b \rangle + \langle b, \mathcal{B} \rangle) = \frac{(2\pi)^2}{\det(\mathcal{A})^{\frac{1}{2}}} \exp(\frac{1}{2}\langle \mathcal{B},\mathcal{A}^{-1} \mathcal{B} \rangle),\\
     	&\int d \chi \exp(-\frac{1}{2}\langle \chi, \mathcal{A} \chi \rangle + \langle \chi, \mathcal{B} \rangle) = \mathrm{Pf}(\mathcal{A}) \exp(\frac{1}{2}\langle \mathcal{B},\mathcal{A}^{-1} \mathcal{B} \rangle).
     \end{align*}
     where $\mathrm{Pf}(\mathcal{A})$ is the Pfaffian of $\mathcal{A}$. % Cite JJ's Geometry and Physics
     The Berezin integral
     \begin{align*}
     	e_{\nabla}^{\mathcal{F}}(t)&:=\frac{1}{(2\pi)^{2m}}\int d\chi db \exp(-t\widetilde{S}_{MQ}) \\
     	&= \frac{1}{(4\pi t)^m} \exp(\frac{t}{4}|\mathcal{F}|^2) \int d\chi \exp(-t(\langle \chi, \nabla \mathcal{F} \rangle - \langle \chi, R_{\nabla} \chi \rangle))\\
     	&= \frac{1}{(4\pi t)^m} \exp(\frac{t}{4}|\mathcal{F}|^2) \mathrm{Pf}(2t R_{\nabla}) \exp(\frac{t^2}{2} \langle \nabla \mathcal{F}, (2t R_{\nabla})^{-1} \nabla \mathcal{F} \rangle) \\
     	&= \frac{1}{(2\pi)^{m}} \mathrm{Pf}(R_{\nabla}) \exp(\frac{t}{4}\left(|\mathcal{F}|^2 + \langle \nabla \mathcal{F}, (R_{\nabla})^{-1} \nabla \mathcal{F} \rangle \right)), \quad t>0,
     \end{align*}
     defines a closed basic $2m$-form on $\mathrm{Fr}(N)$\cite{Mathai1986}. Moreover, $\frac{d}{dt} e_{\nabla}^{\mathcal{F}}(t)$ is exact since $\widetilde{S}_{min}$ is $Q$-exact\footnote{One can also see this by noticing that $|\mathcal{F}|^2 + \langle \nabla \mathcal{F}, ( R_{\nabla})^{-1} \nabla \mathcal{F} \rangle = d \langle \mathcal{F}, \nabla^{-1} \mathcal{F} \rangle$.}, and
     \[
     \lim_{t \rightarrow 0}e_{\nabla}^{\mathcal{F}}(t)= \mathrm{Pf}(\frac{R_{\nabla}}{2\pi}).
     \]
     Thus, $e_{\nabla}^{\mathcal{F}}(t)$ forms a representative of the Euler class of $N$ under the identification $\Omega_{bas}(\mathrm{Fr}(N)) \cong \Omega(N)$. For a transversal $\mathcal{F}$, a proof of the Poincar\'e--Hopf theorem can be obtained by letting $t \rightarrow -\infty$. 
     %also need scalings of the coordinates
     More precisely, we have
    \begin{align}\label{PH}
    	\chi(N) = \lim_{t \rightarrow -\infty} \int_{T[1]N} e_{\nabla}^{\mathcal{F}} (t)
    	= \lim_{t \rightarrow -\infty} \int_N \mathrm{Pf}(\frac{R_{\nabla}}{2\pi}) \exp(\frac{t}{4}|\mathcal{F}|^2)
    	= \sum_{x \in \mathcal{F}^{-1}(0)} \mathrm{ind}_{x}(\mathcal{F}),
    \end{align}
    where $\chi(N)= \int_N \mathrm{Pf}(\frac{R_{\nabla}}{2\pi}) $ is the Euler characteristic of $N$, and $\mathrm{ind}_{x}(\mathcal{F})$ is the degree of the vector field $\mathcal{F}$ at $x$. In the last equality of \eqref{PH}, we use the stationary phase approximation. 
    %(We refer the reader to Mnev and Wernli for more details of the stationary phase approximation in both finite and infinite dimensional settings.).
     
     \subsection{The Atiyah--Jeffrey formula}
     
     The partition function of the minimal CohFT is a finite dimensional Berezin integral. We already see that it can be reduced to the integral
     \[
     \int_{T[1] \mathrm{Fr}(N)} e_{\nabla}^{\mathcal{F}}(t).
     \]
     However, there are two issues: 
     \begin{enumerate}
     	\item $|e_{\nabla}^{\mathcal{F}}(t)|$ goes to infinity when $|\Phi| \rightarrow \infty$ and $t>0$;
     	\item $e_{\nabla}^{\mathcal{F}}(t)$ is not a top form on $\mathrm{Fr}(N)$. 
     \end{enumerate}
    The first issue can be easily solved by replacing $\mathcal{F}$ with $i \mathcal{F}$. To solve the second issue, we need to consider the standard CohFT action functional. In the $0$-dimensional setting, the standard CohFT configuration space is
    \[
    \widetilde{\mathcal{E}}= T[1](\mathfrak{so}(2m)[1]) \times T[1](\mathrm{Fr}(N) \times \mathbb{R}^{2m}[-1] \times \mathfrak{so}(2m)[-2]).
    \]
    Let
    \[
    \widetilde{\mathcal{E}}_{AJ}:= T[1](\mathrm{Fr}(N) \times \mathbb{R}^{2m}[-1] \times \mathfrak{so}(2m)[-2]) .
    \]
    Likewise, one can also define a homomorphism
    \[
    J: C^{\infty}(\widetilde{\mathcal{E}}) \rightarrow C^{\infty}(\widetilde{\mathcal{E}}_{AJ}).
    \]
    The cohomological vector field $Q_J$ takes the same form as before when acting on the fields $\Phi$, $\Psi$, $\chi$, and $b$. For the additional fields $\eta$ and $\lambda$, one can check that
    \[
    Q_J \lambda = \eta - [A_{\nabla}, \lambda], \quad Q_J \eta = -[A_{\nabla}, \eta] + [R_{\nabla}, \lambda].
    \]
    The image of the standard CohFT action functional $\widetilde{S}$ under $J$ is given by
    \[
    J(\widetilde{S}) = \widetilde{S}_{MQ} + \widetilde{S}_{Proj} + s\left( |[R_{\nabla}, \lambda]|^2 - \langle \eta, [R_{\nabla}, \eta] \rangle \right),
    \]
    where
    \[
    \widetilde{S}_{proj} 
    = Q_J  J(\langle \Psi, C(\lambda) \rangle) 
    =  Q_J \langle \Psi + C(A_{\nabla}), C(\lambda)\rangle, 
    \]
    and $C: \mathfrak{so}(2m) \rightarrow T_p \mathrm{Fr}(N)$ is the tangent map of the $\mathrm{SO}(2m)$-action at $p \in \mathrm{Fr}(N)$. It follows that
    \begin{align*}
    	\widetilde{S}_{Proj} &= \langle C (R_{\nabla}), C (\lambda) \rangle - \langle \Psi, C (\eta) \rangle - \langle \Psi, dC(\lambda)(\Psi) \rangle  \\
    	&= \langle C^*C R_{\nabla} - dC^*, \lambda \rangle - \langle C^*, \eta \rangle,
    \end{align*}
    where $C^*: T_p \mathrm{Fr}(N) \rightarrow  \mathfrak{g}$ is the adjoint of $C$. $C^*$ can be viewed as a Lie algebra valued vertical $1$-form on $\mathrm{Fr}(N)$. It is not hard to show that the $2$-form $C^*C R_{\nabla} - dC^*$ is also vertical. Therefore, when being evaluated on horizontal vectors, one can replace $R_{\nabla}$ with $\frac{1}{C^*C}dC^*$. 
    
    On the other hands, the term $s\left( |[R_{\nabla}, \lambda]|^2 - \langle \eta, [R_{\nabla}, \eta] \rangle \right)$ is horizontal. For $m >1$, it will lead to a vanishing path integral because 
    \[
    \dim \mathrm{SO}(2m) = m(2m-1) > 2m = \dim N.
    \]
    Therefore, for a $0$-dimensional CohFT, we need to set $s=0$. To obtain the Atiyah--Jeffrey formula \cite[Equation 2.12.]{Atiyah1990}, we define
    \[
    \widetilde{S}_{AJ}:= \widetilde{S}_{MQ} + i \widetilde{S}_{Proj},
    \]
    where the imaginary unit $i$ is introduced to invoke the Fourier inversion formula. The Berezin integral
    \begin{align*}
    	\gamma_{\nabla}(t)&:=\int d\lambda d\eta \exp(- it \widetilde{S}_{Proj}) \\
    	&=\int d\lambda d\eta \exp(- it \langle C^*CR_{\nabla} - dC^*, \lambda \rangle) \exp(it \langle C^*, \eta \rangle) \\
    	&= i^{\dim \mathrm{SO}(2m)} \int d\lambda \exp(- i \langle R_{\nabla} - \frac{dC^*}{C^*C}, tC^*C \lambda \rangle)\det(tC^*C) \mathrm{vol}_{\pi(p)}
    	%&= \frac{1}{\det(C^*C)}\delta(R_{\nabla} - \frac{dC^*}{C^*C}) 
    \end{align*}
    defines a vertical top form over $\mathrm{Fr}(N)$, where $\mathrm{vol}_{\pi(p)}$ is a volume form on the fiber of $\mathrm{Fr}(N)$ at $\pi(p)$ satisfying
    \[
    \int_{\pi^{-1}(\pi(p))} \mathrm{vol}_{\pi(p)} = 1.
    \]
    $\gamma_{\nabla}(t)$ is independent of $t$ and is called as the projection form in \cite{Cordes1995}. Formally, one can write it as
    \[
    \gamma_{\nabla}(t)= \delta(R_{\nabla} - \frac{dC^*}{C^*C}) \mathrm{vol}_{\pi(p)},
    \]
    where $\delta$ is the Dirac delta function. 
    It follows that
    \begin{align*}
    	\frac{1}{(2\pi)^{\dim N} i^{\dim \mathrm{SO}(2m)}} \int d\chi db d\lambda d\eta \exp(-t \widetilde{S}_{AJ}) = e_{\nabla}^{\mathcal{F}}(t) \wedge \gamma_{\nabla}(t), 
    \end{align*}
    and
    \[
    \int_{T[1]\mathrm{Fr}(N)} e_{\nabla}^{\mathcal{F}}(t) \wedge \gamma_{\nabla}(t) = \int_{T[1]N} e_{\nabla}^{\mathcal{F}}(t).
    \]
    \begin{rmk}\label{sigma}
    	The construction in this section can be directly generalized to the cases of $n$-dimensional sigma model type CohFTs, $n \geq 1$. The configuration spaces of such CohFTs are typically determined by the following data:
    	\begin{itemize}
    		\item $M$ is a $n$-dimensional Riemannian manifold, $N$ is a $2m$-dimensional Riemannian manifold;
    		\item $\mathcal{E} = C^{\infty}(\mathrm{Fr}(M), \mathrm{Fr}(N))_{\mathrm{SO}(n)}$ is the space of $\mathrm{SO}(n)$-invariant smooth maps from $\mathrm{Fr}(M)$ to $\mathrm{Fr}(N)$;
    		\item $\mathcal{E}_{tot} = C^{\infty}_{\mathrm{Lin}}(\mathrm{Fr}(M) \times W, \mathrm{Fr}(N) \times V)_{\mathrm{SO}(n)}$, where $W$ and $V$ are tensor representations of $\mathrm{SO}(n)$ and $\mathrm{SO}(2m)$, respectively;  a $\mathrm{SO}(n)$-invariant smooth map $f=(f_{\mathrm{Fr}(N)}, f_V)$ from $\mathrm{Fr}(M) \times W$ to $\mathrm{Fr}(N) \times V$ is in $\mathcal{E}_{tot}$ if and only if it is linear in the second factors, i.e., 
    		\[
    		f(p, \alpha w) = (f_{\mathrm{Fr}(N)}(p,w), \alpha f_{V}(p,w)), \quad p \in \mathrm{Fr}(M),~ w \in W,~ \alpha \in \mathbb{R};
    		\]
    		the projection map is given by
    		\begin{align*}
    			\pi: \mathcal{E}_{tot}=C^{\infty}_{\mathrm{Lin}}(\mathrm{Fr}(M) \times W, \mathrm{Fr}(N) \times V)_{\mathrm{SO}(n)} &\rightarrow C^{\infty}(\mathrm{Fr}(M) \times \{0\}, \mathrm{Fr}(N))_{\mathrm{SO}(n)} \cong \mathcal{E}   \\
    			f=(f_{\mathrm{Fr}(N)}, f_V) &\mapsto f_{\mathrm{Fr}(N)}|_{\mathrm{Fr}(M) \times \{0\}};
    		\end{align*}
    		\item $\mathcal{G} =C^{\infty}(\mathrm{Fr}(M), \mathrm{SO}(2m))_{\mathrm{SO}(n)}$, for $g \in \mathcal{G}$ and $f \in \mathcal{E}_{tot}$, we define
    		\[
    		g f (p,w)= (f_{\mathrm{Fr}(N)}(p,w)g(p)^{-1}, g(p)f_{V}(p,w));
    		\]
    		%where we use $pg_M$ to denote the right action of $g_M$ on $\mathrm{Fr}(M)$;
    		\item Noting that the $\mathcal{G}$-action is free and
    		\[
    		\mathcal{E}_{tot}/\mathcal{G} \cong C^{\infty}_{\mathrm{Lin}}(\mathrm{Fr}(M) \times_{\mathrm{SO}(n)} W, \mathrm{Fr}(N) \times_{\mathrm{SO}(2m)} V), \quad \mathcal{E}/\mathcal{G} \cong C^{\infty}(M,N);
    		\]
    		hence, a $\mathcal{G}$-equivariant section $\mathcal{F}$ of $\mathcal{E}_{tot}$ is equivalent to a section
    		\[
    		[\mathcal{F}]: C^{\infty}(M,N) \rightarrow C^{\infty}_{\mathrm{Lin}}(\mathrm{Fr}(M) \times_{\mathrm{SO}(n)} W, \mathrm{Fr}(N) \times_{\mathrm{SO}(2m)} V);
    		\]
    		for the ordinary nonlinear sigma model, one can choose $W=\mathbb{R}^{n}$, $V=\mathbb{R}^{2m}$, and 
    		\[
    		[\mathcal{F}]=d\varphi, 
    		\]
    		where $\varphi \in C^{\infty}(M,N)$ and $d\varphi$ is its tangent map,
    		one can also consider the section
    		\[
    		[\mathcal{F}]=\overline{\partial}_J \varphi
    		\]
    		representing the $J$-holomorphic curve equations if $M$ is a Riemann surface and $N$ is a symplectic manifold equipped with a compatible almost complex structure $J$.		
    	\end{itemize}
    	To define the Chern--Weil homomorphism $\mathrm{CW}_{\nabla}$ and the Mathai--Quillen automorphism $T_{\nabla}$ in this infinite dimensional setting, we use the connection $1$-form $A_{\nabla}$ on $\mathrm{Fr}(M) \times \mathcal{E}$ induced by the pullback of the Levi-Civita connection on $\mathrm{Fr}(N)$ under the evaluation map
    	\[
    	\mathrm{ev}: \mathrm{Fr}(M) \times \mathcal{E} \rightarrow \mathrm{Fr}(N).
    	\]
    	It is then straightforward to define the minimal (and standard) CohFTs $(\mathcal{E}_{MQ}, Q_J, \widetilde{S}_{MQ})$ (and $(\mathcal{E}_{AJ}, Q_J, \widetilde{S}_{AJ})$). In particular, the minimal CohFTs of the equations $d\varphi=0$ in dimension $n=1$ and the $J$-holomorphic equations $\overline{\partial}_J \varphi = 0$ in dimension $n=2$ will recover the $N=2$ supersymmetric quantum mechanics and the topological $A$-model \cites{Witten1988b,Baulieu1989}, respectively.   	
    \end{rmk}
    
	%\section{Applications to the generalized Seiberg--Witten equations}
	
	\section{Generalized Seiberg--Witten theory}\label{gsw}
	
	Let $G \subset M_{m \times m}(\mathbb{K})$ be a matrix group containing $-1$, $\mathbb{K}=\mathbb{R}$ or $\mathbb{C}$. The $\mathrm{Spin}^G$ group is defined as
	\[
	\mathrm{Spin}^G(n) := \mathrm{Spin}(n) \times_{\mathbb{Z}_2} G = \mathrm{Spin}(n) \times G/\{(1,1),(-1,-1)\}.
	\]
	There is a short exact sequence
	\[
	\mathrm{Id} \rightarrow \mathbb{Z}_2 \rightarrow \mathrm{Spin}^G(n) \xrightarrow{h} \mathrm{SO}(n) \times G/\mathbb{Z}_2 \rightarrow \mathrm{Id}.
	\]
	The group homomorphism $h$ is defined by sending $[a,g]$ to $(\mathrm{Ad}(a),\pi(g))$, where $\mathrm{Ad}: \mathrm{Spin}(n) \rightarrow \mathrm{SO}(n)$ and $\pi: G \rightarrow G/\mathbb{Z}_2$ are the canonical double coverings.
	A representation $S$ of $\mathrm{Spin}^G(n)$ is said to be a spinor representation if $-1 \in \mathrm{Spin}^G(n)$ acts nontrivially on $S$, and a tensor representation if $-1 \in \mathrm{Spin}^G(n)$ acts as the identity on $S$. Note that every (irreducible) left-/bi-module of $\mathrm{Cl}(n) \otimes_{\mathbb{R}} M_{m\times m}(\mathbb{K})$ can be turned into a spinor/tensor representation of $\mathrm{Spin}^G(n)$. 
	
	Let $M$ be a closed Riemannian manifold. Let $L$ be a principal $G/\mathbb{Z}_2$-bundle over $M$. A $\mathrm{spin}^G$ structure $P$ over $M$ is a principal $\mathrm{Spin}^G(n)$-bundle $P$ together with a bundle morphism
	\[\begin{tikzcd}
		P \arrow{dr} \arrow{r}{\tilde{h}} &  \mathrm{Fr}(M) \times_M L \arrow{d} \\
		& M
	\end{tikzcd}\]
	which restricts fibrewise to the double covering homomorphism $h$.
	%$
	%h:\mathrm{Spin}^G(n) \rightarrow \mathrm{SO}(n) \times G/\mathbb{Z}_2.
	%$
	%When $M$ is $4$-dimensional and $G$ is simply connected, a $\mathrm{spin}^G$ structure always exists, but is not necessarily unique, i.e., it is a structure of $M$ that one needs to specify.
    Let $\slashed{S}$ denote the associated bundle of $P$ determined by a left-/bi-module $S$ of $\mathrm{Cl}(n) \otimes_{\mathbb{R}} M_{m\times m}(\mathbb{K})$. (We call $\slashed{S}$ a generalized Clifford bundle.) The Levi--Civita connection $\nabla$ on $TM$ and a connection $A$ on $L$ induces a $\mathrm{Spin}^G(n)$-connection $\nabla_A$ on $S$. We define the twisted Dirac operator on $\slashed{S}$ via the standard formula
    \[
    \slashed{D}_A \sigma =  (e^{\mu} \otimes 1) \iota_{e_{\mu}} \nabla_A \sigma,
    \]
    where $\sigma \in \Gamma(\slashed{S})$ and $\{e_{\mu}\}_{\mu=1}^n$ is a local orthonormal frame. If $G$ is compact, we can equip $\slashed{S}$ with a $\mathrm{Spin}^G(n)$-invariant bundle metric $\langle \cdot, \cdot\rangle$ that satisfies
    \[
    \langle (e \otimes 1) \sigma_1, (e \otimes 1) \sigma_2 \rangle = \langle \sigma_1, \sigma_2 \rangle
    \]
    for all $\sigma_1, \sigma_2 \in \slashed{S}_x$ and all unit vectors $e \in T_xM$.
    Let $\{\xi_a\}$ be a orthonormal basis of the Lie algebra $\mathrm{Lie}(G/\mathbb{Z}_2)$ of $G/\mathbb{Z}_2$. We define the following quadratic bundle map
    \begin{align}
    	\mu: \slashed{S}  &\rightarrow \Lambda^2 M \otimes \mathrm{ad} L \notag \\
    	\sigma &\mapsto \langle e_{\mu}e_{\nu} \otimes d\pi_{\mathrm{Id}}^{-1}(\xi^a) \sigma, \sigma \rangle (e^{\mu} \wedge e^{\nu}) \otimes \xi_a, \label{mu}
    \end{align}
    where $d\pi_{\mathrm{Id}}: \mathfrak{g} \rightarrow \mathrm{Lie}(G/\mathbb{Z}_2)$ is the tangent map of the double covering homomorphism $\pi: G \rightarrow G/\mathbb{Z}_{2}$ at the identity $\mathrm{Id} \in G$.    
    The GSW equations on $M$ are defined as
    \begin{align}\label{gswo}
    	F_A - \frac{1}{2}\mu(\sigma) = 0,\quad \slashed{D}_A\sigma = 0.
    \end{align}
   From now on, we fix $M$ to be a compact oriented Riemannian $4$-manifold. In such cases, one can replace $\slashed{S}$ with a generalized Clifford bundle $\slashed{S}^{+}$ of positive chirality, and consider instead the GSW equations of the following form
	\begin{align}\label{gsw+}
		(F_A)_{+} - \frac{1}{2}\mu(\sigma) = 0,\quad
		\slashed{D}_A\sigma = 0,
	\end{align}
	where $\sigma$ is a section of $\slashed{S}^{+}$, and $\mu$ is a bundle morphism from $\slashed{S}^{+}$ to $\Lambda^2_{+} M \otimes \mathrm{ad} L$. 
	
	To apply our CohFT construction, we make the following definitions:
	\begin{itemize}
		\item $\mathcal{E} = \mathcal{A}(L) \times \Gamma(\slashed{S}^{+})$ is the product of the affine space of connections on $L$ and the space of sections of $\slashed{S}^{+}$;
		\item $\mathcal{E}_{tot} = \mathcal{E} \times \mathcal{H}$ is a trivial bundle with $\mathcal{H} = \Omega^2_+(M, \mathrm{ad} L) \times \Gamma(\slashed{S}^-)$.
		\item $\mathcal{F}_{\mu}$ is a $\mathcal{G} := \mathrm{Aut}(L)$-equivariant map sending 
		\[
		(A, \sigma) \mapsto (\mathcal{F}_1(A) - \frac{1}{2}\mu(\sigma), \mathcal{F}_2(\sigma)),
		\]
		where 
		$
		\mathcal{F}_1(A)=(F_A)_{+}
		$
		and
		$
		\mathcal{F}_2(\sigma) = \slashed{D}_A \sigma.
		$
	\end{itemize}   
	
	Let's denote $\Phi=(A,\sigma)$, $\Psi=(\psi, \upsilon)$, and, with a slight abuse of notations, $\chi=(\chi, \xi)$, $b=(b,h)$.
	\begin{table}[!h]
		\centering
		\begin{tabular}{ c  c  c  c  c }
			\hline
			deg~2 & deg~1 & deg~0 &  deg~$-1$ & deg~$-2$  \\
			$\phi$ & $\theta$, $\psi$, $\upsilon$ & $A$, $\sigma$, $b$, $h$ & $\chi$, $\xi$, $\eta$ & $\lambda$  \\
			\hline
		\end{tabular}
		\caption{Degrees of the fields.}
	\end{table}
	 For the reader's convenience, we list the actions of $Q$ on the fields below.
	\begin{align*}
		&Q \theta = \phi - \frac{1}{2}[\theta,\theta], \quad Q \phi = -[\theta, \phi],\\
		&Q A = \psi + d_A \theta, \quad Q \psi = -[\theta, \psi] - d_A \phi,\\
		&Q \sigma = \upsilon -\theta \sigma, \quad Q \upsilon = -\theta \upsilon  + \phi \sigma,\\
		&Q \chi = b - [\theta, \chi] +(F_{A})_{+}  - \frac{1}{2} \mu(\sigma), ~
		Q b  = -[\theta, b] + [\phi, \chi] - d_{A}^{+} \psi + \mu(\sigma, \upsilon),\\
		&Q \xi = h - \theta \xi +\slashed{D}_{A} \sigma, ~
		Q h  = -\theta h + \phi \xi - \slashed{D}_{A} \upsilon - \psi \sigma, \\
		&Q \lambda = \eta - [\theta, \lambda], ~
		Q \eta = - [\theta, \eta] + [\phi, \lambda].
	\end{align*}
	where $\mu(\cdot, \cdot)$ is the bilinear form of $\mu(\cdot)$.
	The minimal CohFT action functional of $\mathcal{F}_{\mu}$ is
	\begin{align*}
		\widetilde{S}_{min} &= Q \left( \int_M \langle (\chi,\xi), (b, h) \rangle  \mathrm{vol}_M \right)\\
		&= \int_M \bigg( |b|^2 + \langle (F_A)_{+} - \frac{1}{2} \mu(\sigma), b \rangle + |h|^2 + \langle \slashed{D}_A \sigma, h \rangle \\
		& \qquad  \qquad + \langle \chi, d_A^{+} \psi - \mu(\sigma, \upsilon) \rangle + \langle \xi, \slashed{D}_A\upsilon + \psi \sigma \rangle - \langle \chi, [\phi, \chi] \rangle - \langle \xi, \phi \xi \rangle \bigg) \mathrm{vol}_M,
	\end{align*}
	where we write $\psi\sigma$ to denote the action of $\psi$ on $\sigma$ induced by the Clifford multiplication and the Lie bracket.
	The standard CohFT action functional of $\mathcal{F}_{\mu}$ is
	\begin{align*}
		\widetilde{S} &= \widetilde{S}_{min} + Q \left( \int_M \left(\langle (\psi,\upsilon), (d_A \lambda, -\lambda \sigma) \rangle + \langle \eta, [\phi, \lambda] \rangle \right) \mathrm{vol}_M \right) \\
		&=  \int_M \left(|b|^2 + \langle (F_A)_{+} - \frac{1}{2} \mu(\sigma), b \rangle + |h|^2 + \langle \slashed{D}_A \sigma, h \rangle  - \langle d_A \phi, d_A \lambda \rangle - \langle \phi \sigma, \lambda \sigma \rangle + |[\phi, \lambda]|^2 \right) \mathrm{vol}_M \\
		&+\int_M \bigg(\langle \chi, d_A^{+} \psi \rangle 
		+ \langle \xi, \slashed{D}_A\upsilon \rangle  		
		- \langle \psi, d_A \eta \rangle
		+ \langle \upsilon, \eta \sigma \rangle 
		-\langle \chi, \mu(\sigma, \upsilon) \rangle 
		+ \langle \xi, \psi \sigma \rangle  \\
		& \qquad \qquad - \langle \chi, [\phi, \chi] \rangle 
		- \langle \xi, \phi \xi \rangle 
		+ \langle \psi, [\lambda, \psi]\rangle 
		+ \langle \upsilon, \lambda \upsilon \rangle 
		- \langle \eta,  [\phi, \eta] \rangle  \bigg) \mathrm{vol}_M.
	\end{align*}
    %\subsection{U(2) Seiberg--Witten theory}
    %https://arxiv.org/pdf/math/0203047
    \begin{rmk}
    	Let $G=\mathrm{U}(2)$ and $\slashed{S} = \mathbb{C}^4 \otimes \mathbb{C}^2$ be an irreducible left-module of $\mathrm{Cl}(4) \otimes M_{2 \times 2}(\mathbb{C})$,
    	%a $\mathrm{spin}^G$ structure is called a $\mathrm{spin}^u$ structure. 
    	%Spin^c \times_{U(1)} U(2) \cong Spin^u
    	we have
    	\[
    	\mathrm{U}(2)/\mathbb{Z}_2 \cong \mathrm{U}(1) \times \mathrm{PU}(2) \cong \mathrm{S}^1 \times \mathrm{SO}(3),
    	\]
    	and the connection $A$ on $L$ splits as 
    	\[
    	A = a + A_0.
    	\]
    	Fixing $a$, the resulting GSW equations of $(A_0, \sigma)$ are known as the $\mathrm{SO}(3)$ monopole equations \cite{Pidstrigach1995}. Its moduli space contains both the moduli space of $\mathrm{SO}(3)$ anti-selfdual connections and the moduli space of $\mathrm{U}(1)$ monopoles. The CohFT constructed out of the $\mathrm{SO}(3)$ monopole equations is the twisted $N=2$ $\mathrm{SO}(3)$ supersymmetric Yang--Mills theory with (massless) matter fields \cite[Subsection 5.5.]{Labastida2005}.
    \end{rmk}
    
	\subsection{Donaldson--Witten theory}
	
	Let $G=\mathrm{SU}(2)$ and $\slashed{S}=\{0\}$. $A$ is a connection on a principal $\mathrm{SO}(3)$-bundle $L$. The CohFT action functional $\widetilde{S}$ takes the following form:
	\begin{align*}
		\widetilde{S} &= Q \left(\int_M (\langle \chi, b \rangle + \langle \psi, d_A \lambda \rangle + \langle \eta, [\phi, \lambda] \rangle) \mathrm{vol}_M \right) \\
		&= \int_M \left(|b|^2 + \langle (F_A)_+ , b \rangle - \langle d_A \phi, d_A \lambda \rangle + |[\phi, \lambda]|^2 \right) \mathrm{vol}_M \\
		&+\int_M \left( \langle \chi, d_A^+ \psi \rangle 
		- \langle \chi, [\phi, \chi] \rangle 
		+ \langle \psi, [\lambda, \psi]\rangle 
		- \langle \psi, d_A \eta \rangle  
		- \langle \eta,  [\phi, \eta] \rangle \right) \mathrm{vol}_M.
	\end{align*}
	$\widetilde{S}$ is the CohFT action functional of the twisted pure $N=2$ supersymmetric Yang--Mills theory \cite[Equation 2.13.]{Witten1988}.
	
	\subsection{Seiberg--Witten theory}	
	
	Let $G=\mathrm{U}(1)$. Let $\slashed{S} = \mathbb{C}^4$ be an irreducible left-module of $\mathrm{Cl}(4) \otimes \mathbb{C}$, the GSW equations become the ordinary equations. Applying our CohFT construction yields the twisted $N=2$ $\mathrm{U}(1)$ supersymmetric Yang--Mills theory with (massless) matter fields \cite[Subsection 5.5.]{Labastida2005}.
	
	\subsection{Kapustin--Witten theory}
	
	Let $G=\mathrm{SU}(2)$ and $\slashed{S}= \Lambda T^*M \otimes \mathrm{ad} L$ be an bi-module of $\mathrm{Cl}(4) \otimes M_{2 \times 2}(\mathbb{C})$. In this setting, the twisted Dirac operator can be decomposed as
	\[
	\slashed{D}_A = d_A + d_A^*
	\]
	where $d_A: \Omega^p(\ad L) \rightarrow \Omega^{p+1}(\ad L)$ is the exterior covariant derivative, and $d_A^*$ is the formal adjoint operator of $d_A$. We require $\sigma$ to be a homogeneous $1$-form. One can check that
	\[
	\mu(\sigma) = [\sigma, \sigma].
	\]
	 Consequently, we consider the equivariant map $\mathcal{F}_{\mu}$ sending
	\[
	(A, \sigma) \in \mathcal{A}(L) \times \Omega^1(\mathrm{ad} L) \mapsto ((F_A - \frac{1}{2}[\sigma, \sigma])_+, d_A^- \sigma, d_A^*\sigma) \in \Omega^2_+(\mathrm{ad} L) \oplus \Omega^2_-(\mathrm{ad} L) \oplus \Omega^0(\mathrm{ad} L).
	\]
	With a slight abuse of notations, we redenote the fields $\upsilon$ by $\widetilde{\psi}$, $\xi$ by $(\widetilde{\chi}, \widetilde{\eta})$, and $h$ by $(\widetilde{b}, w)$, where $\widetilde{\chi}$ and $\widetilde{b}$ are anti-selfdual $2$-forms, $\widetilde{\psi}$ is a $1$-form, and $\widetilde{\eta}$ and $w$ are $0$-forms. The scalar supersymmetry $Q$ of the theory takes the following form.
	\begin{align*}
		&Q \theta = \phi - \frac{1}{2}[\theta,\theta], \quad Q \phi = -[\theta, \phi],\\
		&Q A = \psi + d_A \theta, \quad Q \psi = -[\theta, \psi] - d_A \phi,\\
		&Q \sigma =  \widetilde{\psi} -[\theta, \sigma], \quad Q \widetilde{\psi} = -[\theta, \widetilde{\psi}]  + [\phi, \sigma],\\
		&Q \chi = b - [\theta, \chi] + (F_{A}  - \frac{1}{2} [\sigma,\sigma])_+, ~
		Q b  = -[\theta, b] + [\phi, \chi] - d_{A}^+ \psi + [\sigma,  \widetilde{\psi}]_+,\\
		&Q \widetilde{\chi} = \widetilde{b} - [\theta, \widetilde{\chi}] + d_A^- \sigma, ~
		Q \widetilde{b}  = -[\theta, \widetilde{b}] + [\phi, \widetilde{\chi}] - d_A^-  \widetilde{\psi} - [\psi, \sigma]_-, \\
		& Q \widetilde{\eta} = w - [\theta, \widetilde{\eta}] + d_A^* \sigma, ~
		Q w = -[\theta, w] + [\phi, \widetilde{\eta}] - d_A^*  \widetilde{\psi} + \langle [\psi, \sigma] \rangle,\\
		&Q \lambda = \eta - [\theta, \lambda], ~
		Q \eta = - [\theta, \eta] + [\phi, \lambda],
	\end{align*}
	where $\langle [\psi, \sigma] \rangle := \sum_{\mu=1}^4 [\psi_{\mu}, \sigma_{\mu}]$, $\psi = \psi_{\mu} e^{\mu}$, $\sigma = \sigma_{\mu} e^{\mu}$, $\{e^{\mu}\}_{\mu=1}^4$ is a local orthonormal coframe.
	The CohFT action functional is
	\begin{align*}
		\widetilde{S} &= Q \left( \int_M (\langle (\chi, (\widetilde{\chi}, \widetilde{\eta})), (b, (\widetilde{b},w)) \rangle + \langle (\psi, \widetilde{\psi}), (d_A \lambda, -[\lambda ,\sigma]) \rangle + \langle \eta, [\phi, \lambda] \rangle) \mathrm{vol}_M \right). 
	\end{align*}
	$\widetilde{S}$ can be decomposed as $ \widetilde{S} = \widetilde{S}_{even} + \widetilde{S}_{odd}$, where
	\begin{align*}
		\widetilde{S}_{even} 
		&=\int_M \big(|b|^2 + \langle (F_A - \frac{1}{2}[\sigma, \sigma])_+, b \rangle + |\widetilde{b}|^2 + \langle d_A^- \sigma, \widetilde{b} \rangle + |w|^2 + \langle d_A^* \sigma, w \rangle \\  
		& - \langle d_A \phi, d_A \lambda \rangle - \langle [\phi, \sigma], [\lambda, \sigma] \rangle + |[\phi, \lambda]|^2 \big) \mathrm{vol}_M,
	\end{align*}
	and 
	\begin{align*}
		\widetilde{S}_{odd} &= \int_M (\langle \chi, d_A^+ \psi \rangle 
		+ \langle \widetilde{\chi}, d_A^-\widetilde{\psi}  \rangle 
		- \langle \widetilde{\psi}, d_A \widetilde{\eta} \rangle
		- \langle \psi, d_A \eta \rangle 
		- \langle \chi,  [\sigma, \widetilde{\psi}]_+\rangle  
		+ \langle \widetilde{\chi}, [\sigma, \psi]_- \rangle
		+ \langle \psi, [\sigma, \widetilde{\eta}] \rangle \\
		&- \langle \widetilde{\psi}, [\sigma, \eta] \rangle 
		- \langle \chi, [\phi, \chi] \rangle 
		- \langle \widetilde{\chi}, [\phi, \widetilde{\chi}] \rangle 
		- \langle \widetilde{\eta}, [\phi, \widetilde{\eta}] \rangle 
		- \langle \eta,  [\phi, \eta] \rangle
		+ \langle \psi, [\lambda, \psi]\rangle 
		+ \langle \widetilde{\psi}, [\lambda, \widetilde{\psi}] \rangle ) \mathrm{vol}_M.
	\end{align*}
	$\widetilde{S}$ is the CohFT action functional of the twisted pure $N=4$ supersymmetric Yang--Mills theory \cite{Kapustin2007}.
	
	Noting that if there is a bundle isomorphism $\Lambda^2_+T^*M \cong \Lambda^2_- T^*M$,  $\widetilde{S}_{odd}$ is invariant under the following operation:
	\[
	\chi \rightarrow - \widetilde{\chi}, ~\widetilde{\chi} \rightarrow \chi,  \quad \psi \rightarrow -\widetilde{\psi},~ \widetilde{\psi} \mapsto \psi, \quad  \eta \rightarrow - \widetilde{\eta},~ \widetilde{\eta} \mapsto \eta.
	\]
	This is a hint that the CohFT is the complexification of some other CohFT. Let
	\[
	A_c = A + i \sigma, \quad \psi_c = \psi + i \widetilde{\psi}, \quad \chi_c = \chi + i \widetilde{\chi}, \quad \eta_c = \eta + i \widetilde{\eta}, \quad b_c = b + i \widetilde{b}, \quad w_c = iw + [\phi, \lambda].
	\]
	$A_c$ can be interpreted as a $\mathrm{SL}(2,\mathbb{C})$ connection. Its curvature $F_{A_c}$ can be written as
	\[
	F_{A_c} = F_A - \frac{1}{2}[\sigma, \sigma] + i d_A \sigma.
	\]
	Recall that a complex $2$-form $T$ over a $4$-dimensional manifold is said to be selfdual if
	$
	T_{-} = \frac{1}{2}(T - \star \overline{T}) = 0,
	$
	where $\overline{T}$ is the complex conjugate of $T$. It is easy to check that $\chi_c$ and $b_c$ are both complex selfdual $2$-forms. Moreover, the $\mathcal{G}$-equivariant section $\mathcal{F}_{\mu}$ can be reinterpreted as
	\[
	\mathcal{F}_{\mu}(A_c)=((F_{A_c})_+, i \mathrm{Im}(d_{A_c}^* A_c)) \in \Omega^2_+(\mathrm{ad} L; \mathbb{C}) \oplus \Omega^0(\mathrm{ad} L; \mathbb{C}),
	\]
	where we use $\mathrm{Im}(d_{A_c}^* A_c)$ to denote $d_{A_c}^* \mathrm{Im}(A_c) = d_A^* \sigma + i \langle [\sigma, \sigma] \rangle =  d_A^* \sigma$. 
	The expression of $Q$ can be simplified as follows.
	\begin{align*}
		&Q \theta = \phi - \frac{1}{2}[\theta,\theta], \quad Q \phi = -[\theta, \phi],\\
		&Q A_c = \psi_c + d_{A_c} \theta, \quad Q \psi_c = -[\theta, \psi_c] - d_{A_c} \phi,\\
		&Q \chi_c = b_c - [\theta, \chi_c] + (F_{A_c})_+,~
		Q b_c  = -[\theta, b_c] + [\phi, \chi_c] - d_{A_c}^+ \psi_c,\\
		&Q \eta_c = w_c - [\theta, \eta_c] + i \mathrm{Im}(d_{A_c}^* A_c), ~ Q w_c = -[\theta, w_c] + [\phi, \eta_c] - i \mathrm{Im}(d_{A_c}^*  \psi_c),\\
		&Q \lambda = \mathrm{Re}(\eta_c) - [\theta, \lambda],
	\end{align*}
	%and use $\langle [\sigma, \sigma]\rangle = \sum_{\mu=1}^4 [\sigma_{\mu},\sigma_{\mu}]=0$, 	
	The expressions of $\widetilde{S}$, $\widetilde{S}_{even}$, and $\widetilde{S}_{odd}$ can be simplified as follows.
	\begin{align*}
		\widetilde{S} &= Q \left(\int_M (\langle \chi_c, b_c \rangle + \langle \psi_c, d_{A_c} \eta_c \rangle + \langle \eta_c, w_c \rangle) \mathrm{vol}_M \right), \\
		\widetilde{S}_{even} &= \int_M \left(|(b_c, w_c)|^2 + \langle \mathcal{F}_{\mu}(A_c), (b_c, w_c) \rangle   - \langle d_{A_c} \phi, d_{A_c} \lambda \rangle  \right) \mathrm{vol}_M \\
		&=  \int_M \left(|b_c|^2 + \langle F_{A_c}^+, b_c \rangle + |w|^2 + \langle d_A^* \sigma, w \rangle  - \langle d_{A_c} \phi, d_{A_c} \lambda \rangle  + |[\phi, \lambda]|^2 \right) \mathrm{vol}_M,\\
	%\end{align*}
	%The expression of $\widetilde{S}_{odd}$ can also be simplified as follows.
	%\begin{align*}
		\widetilde{S}_{odd} &= \int_M (\langle \chi_c, d_{A_c}^+ \psi_c \rangle 
		- \langle \psi_c, d_{A_c} \eta_c \rangle
		- \langle \chi_c, [\phi, \chi_c] \rangle 
		- \langle \eta_c,  [\phi, \eta_c] \rangle
		+ \langle \psi_c, [\lambda, \psi_c]\rangle) \mathrm{vol}_M.
	\end{align*}
	Therefore, one may think of the Kapustin--Witten theory as the complexification of the Donaldson--Witten theory with an imaginary-gauge fixing condition \cite{Baulieu2009}. This interpretation serves as a CohFT version of the interpretation of \cite{Gagliardo2012}.
	\begin{rmk}
	The configuration space $\widetilde{\mathcal{E}}$ has a $\mathrm{U}(1)$-action, given by
	\[
	g \theta = \theta, ~ g \phi = \phi, ~ g A_c = A_c, ~ g \psi_c = e^{i\varphi} \psi_c, ~ g \chi_c = e^{i\varphi} \chi_c, ~ g b_c = e^{i\varphi} b_c, ~ g \eta_c = e^{i\varphi} \eta_c, ~ g w = w, ~ g \lambda = \lambda, 
	\]
	where $g = e^{i\varphi} \in \mathrm{U}(1)$. Although $\widetilde{S}_{odd}$ is invariant under this $\mathrm{U}(1)$-action, $\widetilde{S}_{even}$ is not since it contains a linear term in $b_c$. However, one can check that
	\[
	\widetilde{S}_{\varphi}:=g^* \widetilde{S}_{\mathcal{F}_{\mu}} = \widetilde{S}_{g \mathcal{F}_{\mu}},
	\]
	where $\widetilde{S}_{\mathcal{F}_{\mu}}$ denotes the CohFT action functional of the PDE represented by $\mathcal{F}_{\mu}$, and 
	\[
    g \mathcal{F}_{\mu} = ((e^{i \varphi} F_{A_c})_+, i \mathrm{Im}(d_{A_c}^* A_c)).
	\]
	Note that $g \mathcal{F}_{\mu}=0$ yields exactly the family of Kapustin--Witten equations:
	\begin{align*}
		\left(\cos(\varphi)(F_A - [\sigma, \sigma]/2) - \sin(\varphi) d_A \sigma\right)_+ &= 0, \\
		\left(\sin(\varphi)(F_A - [\sigma, \sigma]/2) + \cos(\varphi) d_A \sigma\right)_- &= 0, \\
		d_A^* \sigma &= 0.
	\end{align*}	
	In this way, we obtain a family of CohFTs $(\widetilde{\mathcal{E}}, Q_{\varphi}, \widetilde{S}_{\varphi})$, where $Q_{\varphi}:= g Q g^{-1}$.
    \end{rmk}
	%our paper of interest to derived geometer, mathematical gauge theorist and theoretical physicist.
	
	\section{Symmetries and observables}\label{so}
	
	In this section, we discuss the spacetime symmetries and observables of CohFTs in terms of our (minimal) construction. We use the word ``spacetime symmetries" to distinguish from the ``gauge" symmetries of the theory described by the group $\mathcal{G}$. More precisely, we choose $\mathcal{E}$ and $\mathcal{H}$ to be the spaces of sections of some natural bundles $E$ and $H$ over $M$. (We refer the reader to \cite{Fatibene2003} for the notion of a natural bundle.) In such cases, the diffeomorphism group $\mathrm{Diff}(M)$ has a canonical action on $\mathcal{E}_{tot}$, which leads to a $\mathfrak{X}(M)$-action on the configuration space $\widetilde{\mathcal{E}}$ of a CohFT:
	\begin{align*}
		\mathrm{Lie}: \mathfrak{X}(M) &\rightarrow \mathfrak{X}(\widetilde{\mathcal{E}}) \\
		X &\mapsto \mathrm{Lie}_X
	\end{align*}
	where $\mathfrak{X}(\widetilde{\mathcal{E}})$ is the graded Lie superalgebra of derivations of $C^{\infty}(\widetilde{\mathcal{E}})$, $\mathrm{Lie}_X$ is the Lie derivative of $X \in \mathfrak{X}(M)$ on $\widetilde{\mathcal{E}}$ determined by the natural bundle structures of $E$ and $H$.
	\begin{defn}
		An element $X \in  \mathfrak{X}(M)$ is said to be an (infinitesimal) spacetime symmetry of the CohFT $(\widetilde{\mathcal{E}}, Q, \widetilde{S})$ if there exists a degree $-1$ derivation $\iota_{X} \in \mathfrak{X}(\widetilde{\mathcal{E}})$ such that
		\begin{enumerate}
			\item $[Q, \iota_{X}] = \mathrm{Lie}_{X}$;
			\item $\iota_{X} (\widetilde{S}) = 0$.
		\end{enumerate}
	\end{defn}
	By definition, if $X$ is a spacetime symmetry of the theory, then $[Q, \mathrm{Lie}_X]=0$, and
	\[
	\mathrm{Lie}_X(\widetilde{S}) = Q (\iota_X(\widetilde{S})) = 0.
	\]
	%because $\widetilde{S}$ is $Q$-exact. 
	If $X_1$ and $X_2$ are both spacetime symmetries of the theory, then
	\[
    \mathrm{Lie}_{[X_1,X_2]} = [\mathrm{Lie}_{X_1}, \mathrm{Lie}_{X_2}] = [[Q,\iota_{X_1}],  \mathrm{Lie}_{X_2}] =  [Q,[\iota_{X_1},  \mathrm{Lie}_{X_2}]] = [Q, \iota_{[X_1, X_2]}],
	\]
	where $\iota_{[X_1, X_2]}:=[\iota_{X_1}, \mathrm{Lie}_{X_2}]$, and we use the nature bundle assumption to obtain the first equality. (See \cite[Proposition 4.3.6.]{Fatibene2003}.) Moreover, we have
	\[
	\iota_{[X_1,X_2]}(S) = [\iota_{X_1}, \mathrm{Lie}_{X_2}](S) = 0.
	\]
	Thus, $[X_1,X_2]$ is also a spacetime symmetry of the theory.
	We conclude that 
	\begin{prop}
		The spacetime symmetries of the CohFT $(\widetilde{\mathcal{E}}, Q, \widetilde{S})$ form a subalgebra $\mathfrak{s}$ of $\mathfrak{X}(M)$.
	\end{prop}
	Noting that for two spacetime symmetries $X_1$ and $X_2$,
	\[
	[Q,[\iota_{X_1}, \iota_{X_2}]] = [\mathrm{Lie}_{X_1}, \iota_{X_2}] - [\iota_{X_1}, \mathrm{Lie}_{X_2}] = - \iota_{[X_2,X_1]} - \iota_{[X_1,X_2]}=0.
	\]
	but $[\iota_{X_1}, \iota_{X_2}]$ itself does not necessarily vanish. 
	\begin{defn}
		If $[\iota_{X_1}, \iota_{X_2}]=0$ for all $X_1, X_2 \in \mathfrak{s}$, then the $\mathfrak{s}$-action on the configuration space $\widetilde{\mathcal{E}}$ of the CohFT $(\widetilde{\mathcal{E}}, Q, \widetilde{S})$ can be extended to a $\mathfrak{s}_{dR}$-action. In such cases, we say that the CohFT $(\widetilde{\mathcal{E}}, Q, \widetilde{S})$ is de Rham $\mathfrak{s}$-invariant.
	\end{defn}
	
	\begin{exmp}[Vector supersymmetries]
		Let us assume that $M=\mathbb{R}^n$, $\mathfrak{s}$ is the Lie algebra of infinitesimal translations of $M$, i.e., $\mathfrak{s} \cong \mathbb{R}^n$, and the CohFT is de Rham $\mathfrak{s}$-invariant. This is often the case if the CohFT comes from a topological twisting of a supersymmetric field theory on $\mathbb{R}^n$. Fixing a basis $\{\partial_{\mu}\}_{\mu=1}^n$ of $\mathfrak{s}$, we denote $(\partial_{\mu},0) \in \mathfrak{s}_{dR}$ by $K_{\mu}$. ($K_{\mu}$ are called as the vector supersymmetries of the theory by physicists \cites{Sorella1998, Baulieu2005, Piguet2008}.) By definition, we have
		\[
		QK_{\mu} + K_{\mu}Q = \partial_{\mu}, \quad K_{\mu} K_{\nu} + K_{\nu} K_{\mu} = 0, \quad K_{\mu} \partial_{\nu}- \partial_{\nu} K_{\mu}=0.
		\]
		One can also organize $K_{\mu}$ into a single $1$-form symmetry $K:= K_{\mu} dx^{\mu}$. It follows that $QK + KQ = d$, where $d$ is the de Rham differential of $M$. Under some technical assumptions, one can show that $K$ forms a homotopy operator of the $\mathcal{G}^{\star}$ algebra $\Omega(M) \otimes C^{\infty}(\widetilde{\mathcal{E}})$ \cite{Jiang2023a}.\footnote{We refer the reader to \cite{Guillemin2013} for the notion of a $G^{\star}$ algebra, where $G$ is a Lie group.} For a modern mathematical treatment of the vector supersymmetries, we refer the reader to \cites{Elliott2019,Beem2020}.
	\end{exmp}
	
	\begin{exmp}
		Let us assume that $M=\mathbb{CP}^1$, $\mathfrak{s}$ is the Witt algebra, and the CohFT is de Rham $\mathfrak{s}$-invariant. This is often the case if the CohFT comes from a topological twisting of a superconformal field theory. Fixing a basis $\{L_n\}_{n \in \mathbb{Z}}$ of $\mathfrak{s}$, we denote $(L_n, 0) \in \mathfrak{s}_{dR}$ by $G_n$. By definition, we have
		\[
		QG_n +G_nQ=L_n, \quad G_nG_m + G_m G_n =0, \quad G_n L_m - L_m G_n = (m-n)G_{m+n}.
		\]
		The generator $G_{-1}$ together with its complex conjugate can be interpreted as the vector supersymmetries of the theory. For an application of the extended Witt algebra $\mathfrak{s}_{dR}$, we refer the reader to \cite{Losev2018}.
	\end{exmp}
	
	Let us discuss the algebra of classical observables within our minimal construction of CohFTs. Recall that
	\[
		C^{\infty}(\widetilde{\mathcal{E}}_{min}) \cong \mathrm{CE}(\mathrm{Lie}(\mathcal{G})_{dR}; C^{\infty}(T[1](\mathcal{E}_{tot}[-1]))),
	\]
	where both $\mathrm{Lie}(\mathcal{G})$ and $\mathcal{E}_{tot}$ are defined through taking the spaces of sections of some bundles over $M$. Therefore, it makes sense to talk about their restrictions $\mathrm{Lie}(\mathcal{G}|_{U})$ and $\mathcal{E}_{tot}|_U$ to an open subset $U$ of $M$. In other words, we obtain a contravariant functor from the category $\mathbf{Open_M}$ of open subsets in $M$, to the category $\textbf{GraFr\'echetMan}$ of graded Fr\'echet manifolds. On the other hand, taking the Chevalley--Eilenberg complex yields another contravariant functor from $\textbf{GraFr\'echetMan}$ to the category $\mathbf{DGCA}$ of differential graded commutative algebras. Composing these two functors gives us a covariant functor
	\[
	\mathbf{Obs}_{cl}^{\mathcal{F}}: \mathbf{Open_M} \rightarrow \mathbf{DGCA},
	\]
	which assigns to each open subset $U$ of $M$ the differential graded commutative algebra 
	\[
	C^{\infty}(\widetilde{E}_{min}|_{U}) := \mathrm{CE}(\mathrm{Lie}(\mathcal{G}|_U)_{dR}; C^{\infty}(T[1]((\mathcal{E}_{tot}|_U)[-1]))).
	\]
	Under some reasonable locality assumptions on $\mathcal{F}$, one could expect that $\mathbf{Obs}_{cl}^{\mathcal{F}}$ forms a prefactorization algebra on $M$. 
	
	\begin{defn}
		$\mathbf{Obs}_{cl}^{\mathcal{F}}$ is called as the algebra of classical observables of the minimal CohFT of $\mathcal{F}$.
	\end{defn}
	
	A standard approach to defining CohFT observables is through integrating local forms. The key idea is to consider the prefactorization algebra $\mathbf{PreObs}_{cl}^{\mathcal{F}}$ defined by
	\[
	U \mapsto \Omega_{cp}(U) \otimes C^{\infty}(\widetilde{\mathcal{E}}_{min}|_U),
	\]
	where $\Omega_{cp}(U)$ is the de Rham complex of compactly supported differential forms on $U \subset M$. The differential of $ C^{\infty}(\widetilde{E}_{min}|_U)$ is given by the total differential $d + (-1)^{\bullet} Q$, where $(-1)^{\bullet} \omega = (-1)^p \omega$ for $\omega \in \Omega_{cp}^p(U)$, $d$ is the de Rham differential of $\Omega_{cp}(U)$. Let $\mathcal{O}$ be a closed element in $\mathbf{PreObs}_{cl}^{\mathcal{F}}(U)$ of total degree $n$, where $n$ is the dimension of $M$. One can decompose $\mathcal{O}$ as $\mathcal{O} = \sum_{p=0}^n \mathcal{O}^{(p)}$, where $\mathcal{O}^{(p)}$ is of degree $(p,n-p)$ and satisfies
	\begin{align}\label{des}
		Q \mathcal{O}^{(p)} = d \mathcal{O}^{(p-1)}, \quad p=1, \dots, n,
	\end{align}
   and $Q \mathcal{O}^{(0)} = 0$. The equations \eqref{des} are called as the topological descent equations \cite{Witten1988}. Picking a $k$-dimensional closed submanifold $\gamma_k$ inside $U \subset M$, one can define a map
   \begin{align*}
   	\mathrm{ev}_{\gamma_k}|_U:  H^n(\mathbf{PreObs}_{cl}^{\mathcal{F}}(U)) &\rightarrow H^{n-k}(\mathbf{Obs}_{cl}^{\mathcal{F}}(U)) \\
   	[\mathcal{O}] \mapsto [\int_{\gamma_k} \mathcal{O}^{(k)}],
   \end{align*}
   where we use $H^{\bullet}(A)$ to denote the cohomology groups of a differential graded commutative algebra $A$, and use $[a]$ to denote the cohomology class of a closed element $a$ in $A$. It is not hard to see that this map only depends on the homology class of $\gamma_k$.
   \begin{rmk}
   	Physicists refer to observables of the form 
   	$
   	O_{\gamma_k}:=\int_{\gamma_k} \mathcal{O}^{(k)}
   	$
   	as extended observables for $k \geq 1$. Due to their existences, one cannot expect $\mathbf{Obs}_{cl}^{\mathcal{F}}$ to be a factorization algebra (in the usual sense).
   \end{rmk}
	
   \subsection{Vector supersymmetries and observables of the generalized Seiberg--Witten theory}\label{sogsw}
   
   Let us first discuss a standard method of constructing gauge invariant CohFT observables (with $G=\mathrm{SU}(2)$) in the physics literature. Without loss of generality, we only focus on the cases of dimension $n=4$. Let
   \[
   \mathcal{O}^{(0)} = \mathrm{Tr}(\phi^2) \in \Omega^0(M) \otimes  \mathrm{CE}^n(\mathrm{Lie}(\mathcal{G})_{dR}; C^{\infty}(T[1]((\mathcal{E}_{tot})[-1]))),
   \]
   where $\mathrm{Tr}$ is the negative Killing form of the Lie algebra $\mathfrak{su}(2)$.
   $\mathcal{O}^{(0)}$ is $Q$-closed and is of the correct degree $n$ since $\phi$ is of degree $2$. To solve the descent equations \eqref{des}, we define
   \[
   \mathbb{A}(\theta) = \theta + A, \quad \mathbb{F}(\phi) = \phi - \psi + F_A.
   \]
   Recall that 
   \begin{align*}
   	&Q \theta = \phi - \frac{1}{2}[\theta,\theta], \quad Q \phi = -[\theta, \phi],\\
   	&Q A = \psi + d_A \theta, \quad Q \psi = -[\theta, \psi] - d_A \phi.
   \end{align*}
   One can check that \cite{Baulieu1988}
   \[
   \mathbb{F}(\phi) = (d+(-1)^{\bullet}Q) \mathbb{A}(\theta) + \frac{1}{2} [\mathbb{A}(\theta), \mathbb{A}(\theta)], \quad (d+(-1)^{\bullet}Q) \mathbb{F}(\phi) + [\mathbb{A}(\theta), \mathbb{F}(\phi)] = 0.
   \]
   We then define
   \[
   \mathcal{O} = \mathrm{Tr}(\mathbb{F}(\phi)^2).
   \]
   It follows that
   \[
   (d+(-1)^{\bullet}Q)\mathcal{O}  = 2 \mathrm{Tr}\left(((d+(-1)^{\bullet}Q) \mathbb{F}(\phi) + [\mathbb{A}(\theta), \mathbb{F}(\phi)]) \mathbb{F}(\phi)\right) = 0.
   \]
   Expanding $\mathcal{O}$ and integrating its components over the submanifolds of $M$ gives the standard CohFTs observables in the physics literature.
   
   As remarked in Section \ref{gsw}, the GSW theories can be obtained by applying some topological twistings to supersymmetric gauge theories on $\mathbb{R}^4$. This implies that these theories all have vector supersymmetries when $M=\mathbb{R}^4$. Writing down the local expressions for the vector supersymmetry $K_{\mu}$ of a general CohFTs seems not possible. This is because our minimal/standard construction of CohFTs, a priori, does not rely on any supersymmetries. However, in the specific settings of GSW theories, we can indeed find concrete local expressions for $K_{\mu}$, or equivalently, for the $1$-form symmetry $K$. Based on some little experiments, we find that
   %We have
   \begin{align*}
   	&K \theta = A, \quad K \phi = -\psi,  \\
   	&K A = 2\chi, \quad K \psi = 2F_A  - 2b  + \mu(\sigma) - 2(F_A)_+, \\
   	&K \sigma = -e^{\mu} \wedge (e_{\mu} \xi),\quad K \upsilon = e^{\mu} \wedge (e_{\mu} h),\\
   	&K \chi = 0, \quad K b = 3d_A \chi -e^{\mu} \wedge \mu(e_{\mu}\xi, \sigma), \\
   	&K \xi = 0, \quad  K h = -e^{\mu}\wedge \chi_{\mu\nu}(e^{\nu}\sigma),
   \end{align*}
   where $\{e_{\mu}\}_{\mu=1}^4$ is an orthonormal basis of $\mathbb{R}^4$, and we use $e_{\mu} \xi$ to denote the Clifford action of $e_{\mu}$ on $\xi$.   
   One can check that the desired formula
   $
   [Q, K] = d
   $
   holds. For example, we have
   %\[
   %[Q,K] \theta = Q(A) + K(\phi -\frac{1}{2}[\theta, \theta]) = \psi + d_A \theta + (-\psi - [A, \theta]) = d \theta,
   %\]
   %and
  \begin{align*}
  	[Q,K] A &= Q(2 \chi) + K(\psi + d_A \theta) \\
  	&= 2(b - [\theta,\chi] + (F_A)_+ - \frac{1}{2}\mu(\sigma)) + 2F_A  - 2b  + \mu(\sigma) - 2(F_A)_+  - dA - [A,A] \\
  	&=2 F_A - dA - [A,A] = dA,\\
   	[Q,K] \sigma &= Q(-e^{\mu} \wedge (e_{\mu} \xi)) + K(\upsilon - \theta \sigma) \\
   	&= -e^{\mu} \wedge (e_{\mu} (h - \theta \xi + \slashed{D}_A \sigma)) + e^{\mu} \wedge (e_{\mu} h) - A \sigma - e^{\mu} \wedge (e_{\mu} \theta \xi)\\
   	&= -e^{\mu} \wedge (e_{\mu} \slashed{D}_A \sigma) - A \sigma \\
   	&= d_A \sigma - A\sigma = d\sigma,
   \end{align*}
   and 
   \begin{align*}
   	[Q,K] \chi &= K(b - [\theta, \chi] +(F_{A})_{+}  - \frac{1}{2} \mu(\sigma)) \\
   	&=3d_A \chi -e^{\mu} \wedge \mu(e_{\mu}\xi, \sigma) - [A, \chi] - 2 d_A \chi - e^{\mu} \wedge \mu(-e_{\mu}\xi, \sigma) \\
   	&=d \chi,\\
   	[Q,K]\xi &=K(h - \theta \xi + \slashed{D}_A \sigma) \\
   	&=  -e^{\mu}\wedge \chi_{\mu\nu}(e^{\nu}\sigma) - A \xi + e^{\mu} \wedge \chi_{\mu\nu} (e^{\nu} \sigma) - e^{\mu} \wedge \slashed{D}_A (e_{\mu}\xi) \\
   	&= - A \xi + d_A \xi = d \xi.
   \end{align*}
   Moreover, one can check that
   \begin{align}
   	K_{\mu} \left( \int_{\mathbb{R}^4} dx^4 \left(\langle b, \chi \rangle + \langle h, \xi \rangle\right) \right) &= \int_{\mathbb{R}^4} dx^4 \langle D_{\mu} \chi, - \langle  \mu(e_{\mu}\xi, \sigma), \chi \rangle - \langle \chi_{\mu\nu}(e^{\nu}\sigma), \xi \rangle \notag \\
   	&= \int_{\mathbb{R}^4} dx^4 \langle D_{\mu} \chi, \chi \rangle = 0, \label{Kvan}
   \end{align}
   where $D_{\mu}:= \iota_{e_{\mu}} d_A$, and we use the definition \eqref{mu} of $\mu$ to get to the second line. Let $\widetilde{S}_{min}$ be the minimal CohFT action functional. It follows from \eqref{Kvan} that
   \[
   K_{\mu}\widetilde{S}_{min}= K_{\mu} Q \left(\int_{\mathbb{R}^4} d^4 x ( \langle b, \chi \rangle + \langle h, \xi \rangle) \right) = \int_{\mathbb{R}^4} d^4x~ \partial_{\mu} \left(\langle b, \chi \rangle + \langle h, \xi \rangle\right)  = 0.
   \] 
   Here, we use the assumption that all fields are compactly supported on $\mathbb{R}^4$.
	%cite de rham factorization algebra paper
	
   Another wonderful consequence of the existences of $K_{\mu}$ is that they can be used to give an elegant reformulation of the minimal CohFT action functional $\widetilde{S}_{min}$ of Donaldson--Witten theory. To see this, let
   \[
   \theta_K:= \exp(K) \theta =  \sum_{p=0}^4 \frac{1}{p!} K^p \theta, \quad \phi_K := \exp(K) \phi =  \sum_{p=0}^4 \frac{1}{p!} K^p \phi.
   \]
   For Donaldson--Witten theory, a direct computation shows that %\cite[Remark 7.3.]{Jiang2023a}
   \[
   \theta_K= \theta + A + \chi, \quad \phi_K =  \phi - \psi + (b - (F_A)_-) + d_A \chi + \frac{1}{2}[\chi, \chi].
   \]
   $\theta_K$ and $\phi_K$ satisfy relations analogous to those of $\mathbb{A}(\theta)$ and $\mathbb{F}(\phi)$. Moreover, we have \cite{Jiang2023a}
   \begin{align*}
   	\int_M \frac{1}{2}(\exp(K) \mathrm{Tr}(\phi^2))^{top} &= \int_M \mathrm{Tr}(\phi_K^2)^{top} \\
   	& =\int_M \mathrm{Tr}\left(\frac{1}{2}\phi [\chi,\chi] - \psi \wedge d_A \chi + \frac{1}{2} b \wedge b + \frac{1}{2} (F_A)_- \wedge (F_A)_- \right).
   \end{align*}
   Recall that the minimal CohFT action functional of the Donaldson--Witten theory is
   \[
   \widetilde{S}_{min} = \int_M \left(|b|^2 + \langle (F_A)_+ , b \rangle + \langle \chi, d_A^+ \psi \rangle 
   - \langle \chi, [\phi, \chi] \rangle  \right) \mathrm{vol}_M.
   \]
   We have
   \begin{align*}
   	\widetilde{S}_{min} -  \int_M \mathrm{Tr}(\phi_K^2)^{top} = \int_M (\frac{1}{2} |(F_A)_-|^2 + \frac{1}{2} |b|^2 + \langle b, (F_A)_+ \rangle) \mathrm{vol}_M.
   \end{align*}
   ``Integrating out" the auxiliary field $b$, or equivalently, setting $b=-(F_A)_+$, we get 
   \begin{align}\label{hol}
   	\left(\widetilde{S}_{min} -  \int_M \mathrm{Tr}(\phi_K^2)^{top} \right)\bigg|_{b= -(F_A)_+} = \frac{1}{2} \int_M \mathrm{Tr}(F_A \wedge F_A). 
   \end{align}
   \begin{rmk}
   	\eqref{hol} can be explained by the holomorphicity of the $N=2$ supersymmetric gauge theories involving only the $N=2$ vector multiplet. For more details, we refer the reader to \cite[Chapter 4.]{Labastida2005}.
   \end{rmk}
  
   %consider standard CohFT: (dx^{\mu} + \iota_{\partial_{\mu}})K_{\mu}, so we can obtain a Clifford module structure on PreObs? Getzler's proof of index theorem.
   %\[[Q,K]=\slashed{\partial}\]
   
	\section{Quantization program}\label{qp}
	
	So far, we have been focusing mainly on the classical construction of CohFTs. The true power of CohFTs, such as their abilities to produce manifold invariants and quantum cohomologies, only manifests itself in the quantum world. Therefore, it is crucial to consider the quantization of our classical framework. Roughly speaking, the quantization procedure can be carried out in the following four steps.
	\begin{enumerate}[leftmargin=0.52in]
		\item[\textbf{Step 1}] \textbf{Apply the perturbative quantization for a fixed solution $\Phi_{sol} \in \mathrm{Sol}(\mathcal{F})$. }
		
		Due to the existences of various modern mathematical formalisms for perturbative quantum field theories, this will be the easiest step. The key idea is to formulate the CohFT as a cotangent field (or $BF$-like) theory. For such theory, the only quantum contributions are the $1$-loop ones. Moreover, the cotangent field theory formulation of the minimal Donaldson--Witten theory, though not topological, has a strong AKSZ flavor, making it naturally extendable to the cases of manifolds with boundaries. We will discuss this formulation and its relation to the BV-BFV formalism \cites{Cattaneo2014,Cattaneo2018} in detail in another paper.
		
		\item[\textbf{Step 2}] \textbf{Globalize the perturbative series over the entire moduli space $\mathcal{M}(\mathcal{F})=\mathrm{Sol}(\mathcal{F})/\mathcal{G}$.}
		
		This step is more challenging than the first, but the technical difficulties can be addressed. Indeed, powerful globalization techniques developed by Kevin Costello \cite[Chapter 2, Section 13]{Costello2011} and Alberto Cattaneo et al. \cites{Cattaneo2019,Cattaneo2020} are available for this purpose. 
		%and Kevin Costello.
		
		\item[\textbf{Step 3}] \textbf{Integrate the globalized perturbative series over $\mathcal{M}(\mathcal{F})$.} 
		
		This will likely be the most difficult step, as it involves the hard analysis of the compatification problem of the moduli space $\mathcal{M}(\mathcal{F})$.
		%, which our perspective of CohFTs cannot help much. 
		However, 
		%due to the marvelous achievement of the modern mathematical gauge theory community, 
		the compactification theory of several important examples are well known. In particular, the moduli space of the Seiberg--Witten equations is automatically compact. 
		%Converges without compactification.
		%Taubes on Kapustin Witten
		%mention Guillem Cazassus and Wang Donghao's work?
		%mention nonperturbative cohft and derived geometry?
	\end{enumerate}
	
	To summarize, the first step defines a map
	\[
	 \langle \cdot \rangle_{\mathbf{pert}}: \mathbf{Obs}^{\mathcal{F}}_{cl}(M) \rightarrow \mathbb{R}~\text{or}~ \mathbb{C}
	\]
	for each $\Phi_{sol} \in \mathrm{Sol}(\mathcal{F})$ and a collection of tangent vectors at $[\Phi_{sol}] \in \mathcal{M}(\mathcal{F})$. And the second step shows that, for an observable $O$ of degree $k$, $\langle O \rangle_{\mathbf{pert}}$ defines a differential $k$-form over $\mathcal{M}(\mathcal{F})$ in good cases. (In \cite{Witten1988}, Witten suggests that the following formula should holds:
	\[
	\langle O_1 O_2 \rangle_{\mathbf{pert}} = \langle O_1 \rangle_{\mathbf{pert}}  \wedge \langle O_2 \rangle_{\mathbf{pert}}. 
	\]
	We only need to verify this formula at the $1$-loop level.) 
	%For $0 \leq k_i \leq n$, $\sum_{i=1}^l k_i = \dim \mathcal{M}(\mathcal{F})$, 
	The third step then defines a map
	\begin{align*}
		\langle \cdot \rangle_{\mathbf{nonpert}}: ~&\mathbf{Obs}^{\mathcal{F}}_{cl}(M) \rightarrow \mathbb{R}~\text{or}~ \mathbb{C} \\
		& \quad O \mapsto \int_{\mathcal{M}(\mathcal{F})} \langle O \rangle_{\mathbf{pert}}.
	\end{align*}
	$\langle O \rangle_{\mathbf{nonpert}}$ is nontrivial only if the degree of $O$ is $\dim \mathcal{M}(\mathcal{F})$. We write $\langle O \rangle_{\mathbf{nonpert}}^g$ to emphasize its dependence on the geometric data (the Riemannian metric) $g$ of the theory. We still need to show that
	\[
	\frac{d}{dt} \langle O \rangle_{\mathbf{nonpert}}^{g(t)} = 0,
	\]
	where $g(t)$ is a path in the space of geometric data. Here, one will encounter the wall-crossing phenomenon under some specific topological conditions. It would be interesting to compare the results obtained by applying the above quantization procedure, which is essentially built upon the Batalin--Vilkovisky formalism, to the (minimal) Donaldson--Witten theory, with the results in the traditional mathematical literature \cite{Donaldson90}, and with the results obtained by applying the $u$-plane integral techniques \cites{Moore1997, Manschot2023}, which is essentially based on the physical Seiberg--Witten theory \cites{Seiberg/Witten94a,Seiberg/Witten94b}.
	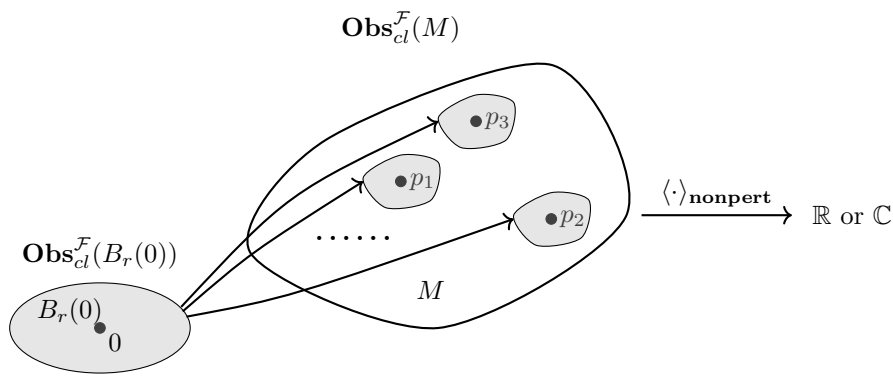
\begin{figure}[ht]
		\centering
		\begin{tikzpicture}
			
			% Draw the manifold as an enlarged irregular shape
			\draw[thick] plot [smooth cycle] coordinates {(3,2.5) (6,3.5) (7,1.5) (4.5,0) (2,1)};
			\node at (4.4,0.5) {$M$};
			
			% Draw the first embedded disk inside the manifold as an irregular shape
			\filldraw[fill=gray!20] plot [smooth cycle] coordinates {(3.5,2) (4,2.3) (4.5,2.2) (4.4,1.7) (3.8,1.6)};
			\filldraw[darkgray] (4,1.95) circle (2pt) node[anchor=west] {$p_1$};
			
			% Draw the second embedded disk inside the manifold as an irregular shape
			\filldraw[fill=gray!20] plot [smooth cycle] coordinates {(5.5,1.5) (6,1.8) (6.5,1.7) (6.4,1.2) (5.8,1.1)};
			\filldraw[darkgray] (6,1.45) circle (2pt) node[anchor=west] {$p_2$};
			
			% Draw the third embedded disk inside the manifold as an irregular shape
			\filldraw[fill=gray!20] plot [smooth cycle] coordinates {(4.5,2.8) (5,3.1) (5.5,3) (5.4,2.5) (4.8,2.4)};
			\filldraw[darkgray] (5,2.75) circle (2pt) node[anchor=west] {$p_3$};
			
			% Draw the elliptic disk outside the manifold
			\filldraw[fill=gray!20] (0,0) ellipse (1.2 and 0.6);
			\filldraw[darkgray] (0,0) circle (2pt);
			\node at (0.2, -0.2) {$0$};
			\node at (-0.4,0.22) {$B_r(0)$};
			
			% Draw the first arrow showing the embedding
			\draw[->, thick] (1.1,0.22) .. controls (2,1) .. (3.5,1.95);
			
			% Draw the second arrow showing the embedding
			\draw[->, thick] (1.15,0.15) .. controls (2.5,0.4) .. (5.5,1.45);
			
			% Draw the third arrow showing the embedding
			\draw[->, thick] (1.08,0.28) .. controls (2.5,1.8) .. (4.5,2.75);
			
			% Draw dots between the arrows
			\node at (3.4,1.2) {\bf{\dots\dots}};
			
			% Draw the arrow for expectation value map
			\draw[->,thick](7.2,1.5) .. controls (8,1.5)..(9.2,1.5);

			% Labels
			\node at (0,1) {$\mathbf{Obs}^{\mathcal{F}}_{cl}(B_r(0))$};
			\node at (4,4) {$\mathbf{Obs}^{\mathcal{F}}_{cl}(M)$};
			\node at (8.2,1.8) {$\langle \bf{\cdot} \rangle_{nonpert}$};
			\node at (10, 1.5) {$\mathbb{R}$ or $\mathbb{C}$};
		\end{tikzpicture}
    \caption{A picture about how to define the quantum product on $\mathbf{Obs}^{\mathcal{F}}_{cl}(B_r(0))$.}		
	\end{figure}
	
	We are now ready to describe the final step.
	\begin{enumerate}[resume,leftmargin=0.52in]
	\item[\textbf{Step 4}] \textbf{Define the quantum cohomology of the theory based on the map $\langle \cdot \rangle_{\mathbf{nonpert}}$ and the embeddings of the open disk $B_r(0)$ into $M$.}
	
	This step will probably yield the most interesting results of the whole quantization program. The key idea can be described as follows. Because of the functoriality of $\mathbf{Obs}^{\mathcal{F}}_{cl}$, one can embed the algebra $\mathbf{Obs}^{\mathcal{F}}_{cl}(B_r(0))$ of classical observables on an open ball $B_r(0)$ of radius $r$ into the the algebra $\mathbf{Obs}^{\mathcal{F}}_{cl}(M)$ of classical observables on $M$. We consider the images of observables in $\mathbf{Obs}^{\mathcal{F}}_{cl}(B_r(0))$ under several such embeddings that are mutally disjoint, multiply them together inside $\mathbf{Obs}^{\mathcal{F}}_{cl}(M)$, and then apply the map $\langle \cdot \rangle_{\mathbf{nonpert}}$. In this way, one will obtain a deformation of the classical product on $\mathbf{Obs}^{\mathcal{F}}_{cl}(B_r(0))$ -- the quantum product. To get rid of the dependence on $r$, we need to show that $\mathbf{Obs}^{\mathcal{F}}_{cl}$ can be equipped with a $\mathbb{R}^n_{dR}$-action\footnote{This, in particular, holds for the generalized Seiberg Witten theories. See Subsection \ref{sogsw}.} and invoke the Corollary 2.30 of \cite{Elliott2019}.
	\end{enumerate}
	
	There exists a vast literature on quantum cohomologies in the settings of $2$-dimensional conformal field theories. In particular, Kontsevich and Manin proposed a beautiful axiomatic framework for quantum cohomologies, which they also refer to as a cohomological field theory \cite{Kontsevich1994}. It would be very interesting to compare their notion and our notion of CohFTs (which is essentially due to Witten \cite{Witten1991}) in the context of $2$-dimensional conformal field theories. Moreover, our quantization program can also be used to define quantum cohomologies for $4$-dimensional gauge theories, and in particular, for the generalized Seiberg--Witten theories. In such cases, the cohomology groups of the algebra $\mathbf{Obs}^{\mathcal{F}}_{cl}(B_r(0))$ of classical observables on $B_r(0)$ might not be as interesting as the $2$-dimensional cases. However, the quantum product obtained in Step 4 above is a product at the cochain complex level. Therefore, one may still hope that it encodes richer and finer information about the geometry of $4$-manifolds.
	
	\section{Conclusions and discussions}\label{cd}
	
	We have proposed a mathematical framework for cohomological field theories. The framework relies solely on the data of a nonlinear PDE, which allows one to use it to obtain a uniform description of various theories in the physics literature. 
	
	In this paper, our primary focus is the application of the framework to the generalized Seiberg--Witten equations in dimension $4$. However, as highlighted in Remark \ref{sigma}, our framework can also be used to provide a uniform description of the sigma model type theories. In particular, it would be interesting to consider its application to the Euler--Lagrange equations of the Poisson sigma model \cites{Ikeda1994,SCHALLER1994}. For further applications, there is no reason for one to restrict his or her attention to the ordinary PDEs -- he or she may also consider supersymmetric nonlinear PDEs, such as the super J-holomorphic curve equations, which were introduced in \cite{Jost2017} with the aim to define a theory of super quantum cohomology \cite{Keßler2021}. (For that purpose, we need to require $\mathcal{E}_{tot}$ to be a super Fr\'echet vector bundle and require $\mathcal{G}$ to be a Lie supergroup. The resulted CohFT will possess a $\mathbb{Z}_2 \times \mathbb{Z}$-grading instead of a $\mathbb{Z}$-grading.)
	
	There are several long-term future research directions stemming from our work, as already outlined in Section \ref{qp}. Here, we describe another interesting but challenging direction, also related to the quantization problem of CohFTs. 
	
	Recall that in our quantization program, we use a hybrid form of quantization: we first apply the perturbative BV quantization for a fixed solution in $\mathrm{Sol}(\mathcal{F})$, and then apply the nonperturbative quantization by integrating the perturbative result over a nice moduli space $\mathcal{M}(\mathcal{F})$. The reason we do this is that we want to stay close to and benefit from physicists' intuition. However, as we already hinted in Subsection \ref{minCohFT}, from a mathematical point of view, the infinite dimensional equivariant de Rham cohomology theory involved in our minimal construction of CohFTs is actually a very bad cohomology theory for the space $\mathrm{Sol}(\mathcal{F})$ because it assumes implicitly the smoothness of $\mathrm{Sol}(\mathcal{F})$. One may try to reformulate another cohomology theory for $\mathrm{Sol}(\mathcal{F})$ based on the simplicial de Rham complex of the classifying space $B\mathcal{G}$ of the infinite dimensional Lie group $\mathcal{G}$, viewed as a simplicial manifold. By doing so, one can still get a prefactorization algebra $\mathbf{Obs}_{cl}^{\mathcal{F}}$ of classical observables, but he or she will lose two of the most important field-theoretic ingredients that are needed for the purpose of applying perturbative quantization: the CohFT action functional and its scalar supersymmetry. Therefore, the quantization theory for the newly constructed $\mathbf{Obs}_{cl}^{\mathcal{F}}$, if it exists, is expected to be nonperturbative by the very nature of $\mathbf{Obs}_{cl}^{\mathcal{F}}$. (To tackle the problem of defining such quantization theory, one can and should start with the $0$-dimensional toy model.)

	\end{document}